\documentclass[sigconf, final]{acmart}

\AtBeginDocument{%
  }

\setcopyright{acmlicensed}
\copyrightyear{2024}
\acmYear{2024}
\acmDOI{XXXXXXX.XXXXXXX}

\acmConference[Conference acronym ’XX]{Make sure to enter the correct
  conference title from your rights confirmation emai}{June 03--05, 2018}{Woodstock, NY}
\acmISBN{978-1-4503-XXXX-X/18/06}
\usepackage{makecell}
\usepackage{romannum}
\usepackage{multirow}
\usepackage{caption}
\usepackage{graphicx,subfig}

\begin{document}

\title[To Shelter or Not To Shelter]{To Shelter or Not To Shelter: Exploring the Influence of Different Modalities in Virtual Reality on Individuals' Tornado Mitigation Behaviors}

%% Exploring the Design of Different Modalities in Virtual Reality for Improving Individuals' Risk Perception and Mitigation Behaviors Related to Tornado Hazards


\author{Jiuyi Xu}
\authornote{Both authors contributed equally to this research.}
\affiliation{%
  \institution{Colorado School of Mines}
  \city{Golden}
  \state{Colorado}
  \country{USA}
}
\email{jiuyi_xu@mines.edu}

\author{Tolulope Sanni}
\authornotemark[1]
\affiliation{%
  \institution{California State University, Fresno}
  \city{Fresno}
  \state{California}
  \country{USA}
}
\email{tosanni@mail.fresnostate.edu}

\author{Ziming Liu}
\affiliation{%
  \institution{Colorado School of Mines}
  \city{Golden}
  \state{Colorado}
  \country{USA}
}
\email{ziming_liu@mines.edu}

\author{Yang Ye}
\affiliation{%
  \institution{Northeastern University}
  \city{Boston}
  \state{Massachusetts}
  \country{USA}}
  \email{y.ye@northeastern.edu}
  
\author{Jiyoung Lee}

\affiliation{%
  \institution{Sungkyunkwan University}
  \city{Seoul}
  \country{Republic of Korea}}
  \email{jiyoung.lee@g.skku.edu}

\author{Wei Song}

\affiliation{%
  \institution{University of Alabama}
  \city{Tuscaloosa}
  \state{Alabama}
  \country{USA}
}
  \email{wsong@eng.ua.edu}

\author{Yangming Shi}
\affiliation{%
  \institution{Colorado School of Mines}
  \city{Golden}
  \state{Colorado}
  \country{USA}
}
  \email{yangming.shi@mines.edu}

\renewcommand{\shortauthors}{Xu et al.}

\begin{abstract}
Timely and adequate risk communication before natural hazards can reduce losses from extreme weather events and provide more resilient disaster preparedness. However, existing natural hazard risk communications have been abstract, ineffective, not immersive, and sometimes counterproductive. The implementation of virtual reality (VR) for natural hazard risk communication presents a promising alternative to the existing risk communication system by offering immersive and engaging experiences. However, it is still unknown how different modalities in VR could affect individuals’ mitigation behaviors related to incoming natural hazards. In addition, it is also not clear how the repetitive risk communication of different modalities in the VR system leads to the effect of risk habituation. To fill the knowledge gap, we developed a VR system with a tornado risk communication scenario and conducted a mixed-design human subject experiment (N = 24). We comprehensively investigated our research using both quantitative and qualitative results.

\end{abstract}




% \received{12 September 2024}

\maketitle

\section{INTRODUCTION}
Due to the impact of climate change, the frequency and severity of natural hazards, such as hurricanes, flooding, heat waves, winter storms, wildfires, tornadoes, and other extreme weather events, are increasing, affecting communities across the United States (U.S.)\cite{banholzer2014impact}. Between 2021 and 2023, climate-related natural hazards caused 1,460 fatalities and more than \$400 billion in GDP losses, and other immeasurable damages in the U.S.\cite{NAP13457,NCEI2023,chiba2017climate}. Unfortunately, current research suggests that natural disaster losses are expected to increase in the future, raising concerns that these events will become more common and more destructive due to climate change\cite{peterson2008monitoring}. 

One effective approach to mitigating natural hazards is to deliver effective disaster warning messages, a preemptive measure to protect individuals from potential hazards. In fact, the government alerts individuals in advance, urging them to prepare for an impending event even before it is officially forecasted by weather service agencies and emergency management agencies\cite{keeney2012multi,kim2010communication}. These disaster warning systems delivered through various communication channels (e.g. text messages, emergency alerts, radio/television broadcasts) are designed to motivate individuals to take precautionary measures\cite{national2018emergency,ortiz2023developing}. However, current natural-hazard risk communication delivered through these channels often conveys abstract and generalized information\cite{ortiz2023developing}, making it difficult for individuals to fully imagine, visualize, and engage with the messages. For example, a generic text alerting message stating that "Severe weather is expected" may fail to adequately inform individuals, especially those without prior experience with natural hazards, due to their lack of understanding about the specific consequences of an approaching tornado. Höppner et al.\cite{hoppner2010risk} underscored the difficulty in risk communication to convey information about an event, especially when no specific guidance is provided, leaving recipients feeling helpless until the disaster occurs. The inability of natural hazard risk communications to deliver information that prompts adequate and effective protective actions can severely hinder efforts to mitigate the impact of such natural hazards. 

Then, how can warning messages be delivered in a more immersive and effective way? Recently, research on the effectiveness of immersive technologies in risk communication has yielded promising findings. Studies indicate using Virtual Reality (VR) to deliver risk-related messages has been shown to increase engagement with the core content, helping individuals better equip themselves for hazard preparedness\cite{ogie2018disaster,mol2022after}. Empirical research has supported the potential of implementing VR for the communication of risks from natural hazards. For example, the study  \cite{mitsuhara2020comparative} has suggested that audiovisual effects, particularly those delivered through VR, can help maintain a heightened sense of fear about tornadoes. However, despite these advancements, there is still a significant human-computer interaction (HCI) knowledge gap in research on the impact of different modalities, such as text, audio, and haptics feedback, in VR systems on individuals' mitigation behavior and risk habituation. Risk habituation refers to the phenomenon where individuals become desensitized to certain risks over time due to repeated exposure, leading to a diminished response or reduced perceived seriousness of the risk\cite{slovic1987perception}. This behavior can result in individuals taking fewer precautions or failing to act appropriately in potentially dangerous situations, especially when the perceived risk no longer triggers the expected level of concern or attention. Addressing this gap is particularly crucial for two primary reasons: (a) different modalities can elicit varying psychological responses to risk messages, so identifying the most effective combinations and understanding their interplay can enhance our theoretical understanding of how to optimize VR risk warning messages; and (b) from a practical standpoint, understanding these effects can guide the development of more effective VR-based communication tools that improve public preparedness and response during natural hazards. That said, rather than generalizing that VR is more effective, it is now more important than ever to unpack the specific elements that consist of VR risk messages. Hence, the primary purpose of this study is twofold: 
\begin{itemize}
\item \textbf{RQ1} - Does the use of different modalities in the VR natural hazard risk communication system affect individuals' mitigation behavior differently?
\item \textbf{RQ2} - How does the risk habituation effect vary across different modalities in the VR system? 
\end{itemize}

To answer these research questions, we developed a VR system that simulated a virtual tornado warning scenario and conducted a human-subject experiment with 24 participants that included a mixed experimental design. It aims to examine how different risk communication modalities influence individuals' mitigation behavior and risk habituation. Six different sequences of three tornado risk communication modalities (i.e., text, audio, and mix of text and haptics) will be evenly and randomly assigned for the 24 participants. During the experiment, participants will wear our devices and follow the assigned modality sequence to complete three treatments of 22 one-minute trials of a VR game in different modalities. In each trial, participants will receive a risk alert, but a tornado will only actually appear in three of those trials. We will collect the participants' aggregated shelter time and first shelter entry time data in each trial as indicators of their task performance. Additionally, after each trial, we will gather data on participants' risk perception, behavioral proactivity, and trust in the alert for analysis. After each modality treatment, participants will be asked to complete the NASA-TLX, VR Sickness, and VR Presence questionnaires to evaluate our VR system. We also collected participants' eye-tracking data in an attempt to explore potential connections between the cognitive load and their mitigation behaviors and between the cognitive load and their risk habituation.

We propose the following contributions:
\begin{itemize}
\item A novel immersive approach to improve understanding and response to natural hazard risk-related information across individuals.
\item A comprehensive mixed experiment that evaluates the effect of different modalities on individuals' mitigation behavior and risk habituation.
\item An insight for researchers interested in leveraging immersive technologies to improve public awareness and response to natural hazards.
\end{itemize}

\section{RELATED WORK}
\subsection{Existing Natural Hazard Risk Communication Methods}
Risk communication has emerged as a distinct and increasingly important field, particularly gaining significant attention in the late 20th century. Research has consistently shown that effective risk communication is crucial for hazard management and preparedness, as it plays a key role in mitigating the impacts of natural hazards. Risk communication about incoming natural hazards can prompt and encourage proactive behaviors that save lives\cite{mayhorn2014warning}, minimize potential property damage\cite{ortiz2023developing}, and prevent psychological or emotional distress\cite{hoppner2010risk}.

However, there are ongoing concerns about the potentially conflicting nature of risk communication messages. Yamori\cite{yamori2020disaster} argued that subtle differences in the intended effects often clash when messages are overly generalized, leading to varied responses among different audiences. Unfortunately, many studies have highlighted that generalized media risk communication during an impending natural hazard event may be accompanied by distrust, lack of engagement, and other factors prompting individuals to take actions that could further jeopardize lives and property\cite{lundgren2018risk,wachinger2013risk}. Feldman et al.\cite{feldman2016communicating} find that older populations tend to rely on traditional media, while younger populations show a preference for social media, although the latter is less trusted for flood-related information. This finding emphasizes the need for diverse communication strategies that accommodate different age groups to effectively convey flood risks. The most effective format for risk communication is still a topic of debate, with some studies suggesting that it may be necessary to include persuasive elements to encourage recipients to take protective actions or modify their behaviors\cite{lindell2012protective}.

\subsubsection{Multi-Channel Communication}
Recent studies highlight the importance of integrated communication strategies that combine traditional media, social media, and direct messaging systems through mobile phones\cite{paton2007preparing,sutton2008backchannels,susmayadi2014sustainable}. This multi-channel approach ensures broader reach and more effective dissemination of information, particularly in diverse and vulnerable communities\cite{backfried2016general}. Particularly, the integration of social media has been highlighted for its role in providing real-time updates and actively engaging communities during hazards, enabling prompt information sharing and fostering a sense of connection among those affected \cite{houston2015social,austin2016social}.

\subsubsection{Community Engagement and Education}
In addition to leveraging various communication channels, it is crucial to engage communities through education and participatory methods to enhance hazard preparedness\cite{johnston2022engaging,rahmayati2017understanding},. Programs involving community members in educational campaigns effectively improve knowledge and readiness for hazard events\cite{paton2013preparing}. These programs focus on educating the public through immersive methods, which not only increase awareness, but also build trust and cooperation between the community and hazard management agencies. Research highlights that these methods lead to better outcomes during actual hazards by empowering individuals with the knowledge they need to respond effectively\cite{herbert2023improving}.

\subsubsection{Early Warning Systems}
Implementing early warning systems, a preemptive measure taken before disasters strike, is a key strategy in risk communication. These systems are designed to provide timely alerts to the public, allowing preventive actions, such as seeking shelter, to be taken before a hazard strikes. Recent advancements have incorporated Artificial Intelligence (AI), including machine learning, to improve the accuracy and timeliness of warnings. For example, Jiang et al.\cite{jiang2022scientometric} discussed how AI, particularly machine learning, is increasingly applied to geohazard prediction, with advancements in identifying patterns that can predict earthquakes and landslides. Similarly, Lin et al.\cite{lin2021early} further demonstrated the application of deep learning in early earthquake detection by analyzing crustal deformation patterns, contributing to more timely and accurate earthquake warnings. These studies collectively underscore the value of AI in processing large datasets to provide earlier, more reliable alerts, significantly improving disaster preparedness and mitigation efforts. Studies have shown that such systems, when combined with clear communication protocols, significantly reduce the impact of natural hazards by enabling timely evacuations and other protective measures\cite{kelman2020disaster, mcglade2019global}, as further evidenced by research demonstrating a correlation between timely warnings and reduced casualties\cite{basher2006global}.

\subsection{Different Modalities of Risk Communication}
Researchers have extensively examined the effects of using various modalities to understand the impact of messages. The general conclusion from these researches is messages incorporating multiple modalities—such as a combination of text and visuals—tend to exert a more decisive influence than those relying on a single modality\cite{lee2023emotional,lee2022something,wu2020exploring}. Multimodality is particularly significant in immersive media like VR. Some scholars have combined narration, voice commands, and haptic feedback to enhance user awareness of environmental issues, including air and noise pollution\cite{katsiokalis2023gonature}. VR's ability to integrate multimodal sensory information facilitates immersive experiences that exceed the engagement levels of traditional media, which often depend on singular modalities\cite{martin2022multimodality}.

Despite the inherent multimodality of immersive media environments, there is still a limited understanding of how specific modalities exert more influence than others within these contexts. Although these environments inherently support multiple modalities, the delivery of information can still be dominated by particular sensory inputs. Previous research has thrown light on the differential impact of these modalities when used in isolation. For instance, in scenarios where communication is based solely on audio or visual cues, Spence et al.\cite{spence2012colavita} consistently identified the "visual dominance effect," demonstrating that visual information generally exerts a stronger influence than audio, even when the visual component is less prominent. The adage "a picture is worth a thousand words" reflects the common perception that visuals are often more powerful than text, as they offer a direct representation of reality, in contrast to the abstract nature of textual information. This phenomenon is encapsulated by the realism heuristic\cite{sundar2018heuristic}, which suggests that visuals trigger cognitive shortcuts, enabling individuals to process information more quickly than text, thereby enhancing persuasive effects.

Additionally, although haptic feedback has been relatively understudied, its ability to replicate sensations and textures has made it a valuable component of VR, offering a medium that closely mimics real-world experiences\cite{israr2014feel,Preechayasomboon2021HapletsFW}. A study by Gibbs et al.\cite{GIBBS2022102717} found that combining multiple modalities, including haptics, resulted in a stronger sense of presence than visual or haptic feedback alone. Haptic feedback can enhance the realism and sense of immersion in virtual environments, potentially leading to more effective perception and behavioral changes. In natural natural risk communication, providing quality information in various formats is crucial to guide and prepare recipients for protective actions adequately. However, some studies have cautioned that excessive information could diminish its effectiveness\cite{GAGLIARDI2023106141}. By extending established conceptual frameworks to immersive media settings in risk communication, this exploration seeks to identify which modalities—such as text, visuals, audio, and haptics—function as dominant channels for effectively conveying alert messages.

\subsection{VR Technologies in Risk Communication}
VR is a technology that creates realistic, immersive environments, allowing users to feel present within these spaces by enabling interaction through hand movements, voice, and body gestures\cite{burdea2003vr,SIMPSON2022101764}. VR simulates a three-dimensional(3D) environment that users can engage with using specialized hardware, such as headsets and controllers\cite{ritter2022three}. Research has demonstrated the benefits of using immersive techniques in enhancing hurricane risk communication. For example, Irby et al.\cite{irby2009improving} compared visual representations in a simulated hurricane scenario to traditional presentation methods, finding that VR significantly improved understanding of hurricane risks while offering the advantage of visualizing these risks within a controlled environment. Furthermore, VR can simulate evacuation scenarios, allowing users to practice evacuation plans safely and effectively, while also being employed to develop effective training programs for professionals in hazard preparedness \cite{SHI2021105231}. A systematic review by Hsu et al.\cite{5e941ffe32bc4cdebb72fcf9cdf55598} highlighted the advantages of VR over traditional training methods, concluding that VR-based training for emergency responders could significantly enhance hazard preparedness.

In recent years, research on hazard management has increasingly focused on the use of VR for immersive hazard visualization. Studies indicate that VR can enhance users' understanding of natural hazards and improve decision-making by accurately representing objects' positions within a 3D environment \cite{irby2009improving}. For instance, Sermet and Demir\cite{sermet2018flood} developed "FloodVR," an educational tool that utilizes historical meteorological data and immersive 3D visualizations to depict flooding hazards vividly. Oyshi et al.\cite{oyshi2022floodvis} investigated the impact of VR on decision-making by facilitating the visualization of flood projection data, exploring whether the sense of presence in VR influences the decision-making process in flood risk reduction measures. Their findings revealed that participants who experienced flood impacts through VR had a deeper understanding of the consequences of extreme flood events, suggesting that VR is a valuable tool for visualizing weather- and climate-related hazards and risks. Similarly, Molan et al.\cite{molan2022why} used VR to study participants' evacuation behaviors in response to wildfire hazards, concluding that VR can effectively influence individuals' decisions regarding protective actions. These previous studies have underscored the benefits of visualization and have developed visual prototypes to emphasize the potential of VR in natural hazard risk communication. However, none of these studies have systematically explored strategies for using VR as a supplementary tool in risk communication for natural hazards, nor have they positioned VR as the primary platform for communicating risks while evaluating its impact on individuals' mitigation behaviors and risk habituation with different modalities. 

\subsection{Mitigation behavior and Risk Habituation}
Individuals’ mitigation behavior in response to natural hazards is shaped by various cognitive, social, and environmental factors. Research highlights that risk perception is central to motivating protective actions, such as retrofitting homes, evacuating, or purchasing insurance, as individuals who perceive greater risks tend to take more precautions\cite{bubeck2012review}. However, repeated exposure to natural hazards without significant consequences can lead to risk habituation, where the perceived threat diminishes over time, causing complacency or reduced action\cite{wachinger2013risk}. Social factors, including community norms and peer behaviors, also play a role in shaping habituation, as collective inaction may discourage individual mitigation efforts\cite{poussin2014factors}. Furthermore, effective risk communication strategies are critical in mitigating the effects of habituation by maintaining risk awareness and encouraging proactive behavior. Messages tailored to specific audiences and focused on personal relevance have been found to increase preparedness and counteract the effects of habituation\cite{poussin2014factors}. Therefore, understanding the relationship between mitigation behavior, risk habituation, and communication is essential for developing interventions that foster sustained mitigation behaviors in response to natural hazards.

In the context of VR, the relationship among different modalities, such as text, audio, and haptics, plays a critical role in shaping user behavior, particularly in terms of risk perception, mitigation, and habituation. Multimodal interactions can influence how individuals perceive and respond to potentially risky environments in VR, with some modalities leading to heightened awareness of risks and others facilitating faster habituation. Risk habituation occurs when repeated exposure to the same stimuli results in a diminished response to perceived threats, potentially lowering a user's engagement in risk mitigation behaviors. This habituation is often studied in VR because of its immersive nature, which can simulate real-world risk scenarios. The diminishing response to risk can be due to factors like overexposure to certain modalities or the predictable nature of the risk presented, making users feel overly familiar and less cautious. According to research, this phenomenon is critical as it has implications for training and safety protocols in virtual environments\cite{meehan2002physiological,riva2019virtual}. Specifically, studies have shown that repeated exposure in VR can lead to decreased physiological and psychological responses to risk, reinforcing the importance of managing habituation during training\cite{bailenson2003interpersonal}.

\section{METHOD}
We developed a VR risk communication system designed to simulate tornado scenarios. Based on this VR system, we conducted a mixed (within-participants and between-participants) experiment to explore how different modalities affect individuals' mitigation behavior and risk habituation. The study was approved by the Institutional Review Board (IRB) of the University of Alabama.

\subsection{The overview of the VR system}
\subsubsection{Apparatus}
Our VR system is developed with the Unity3D platform (2019.4.30f1)\cite{unity} and runs on the HTC VIVE Pro Eye enabled head-mounted display (HMD)\cite{HTC}. Our setup incorporated a TactSuit X40\cite{X40} featuring 40 feedback points, ensuring a highly immersive experience with precise environmental feedback. The system was powered by a desktop equipped with an Nvidia GeForce RTX 4090 graphics card, an Intel Core i9-11900k processor clocked at 5.3 GHz, and 64GB of RAM. The VR controller comprises multiple integrated modules designed to enhance the realism of the virtual experience. It features sensory simulation for auditory and visual input, as well as haptic feedback that vibrates in response to sound cues. Eye-tracking technology is employed to monitor user focus and identify points of interest based on our work\cite{shi2020impact2}. Additionally, a locomotion module allows users to navigate the virtual environment by swinging their arms, with the speed of movement directly proportional to the frequency of each swing.

\subsubsection{Virtual Environment Setup}
\begin{figure}[h]
  \centering
  \includegraphics[width=\linewidth]{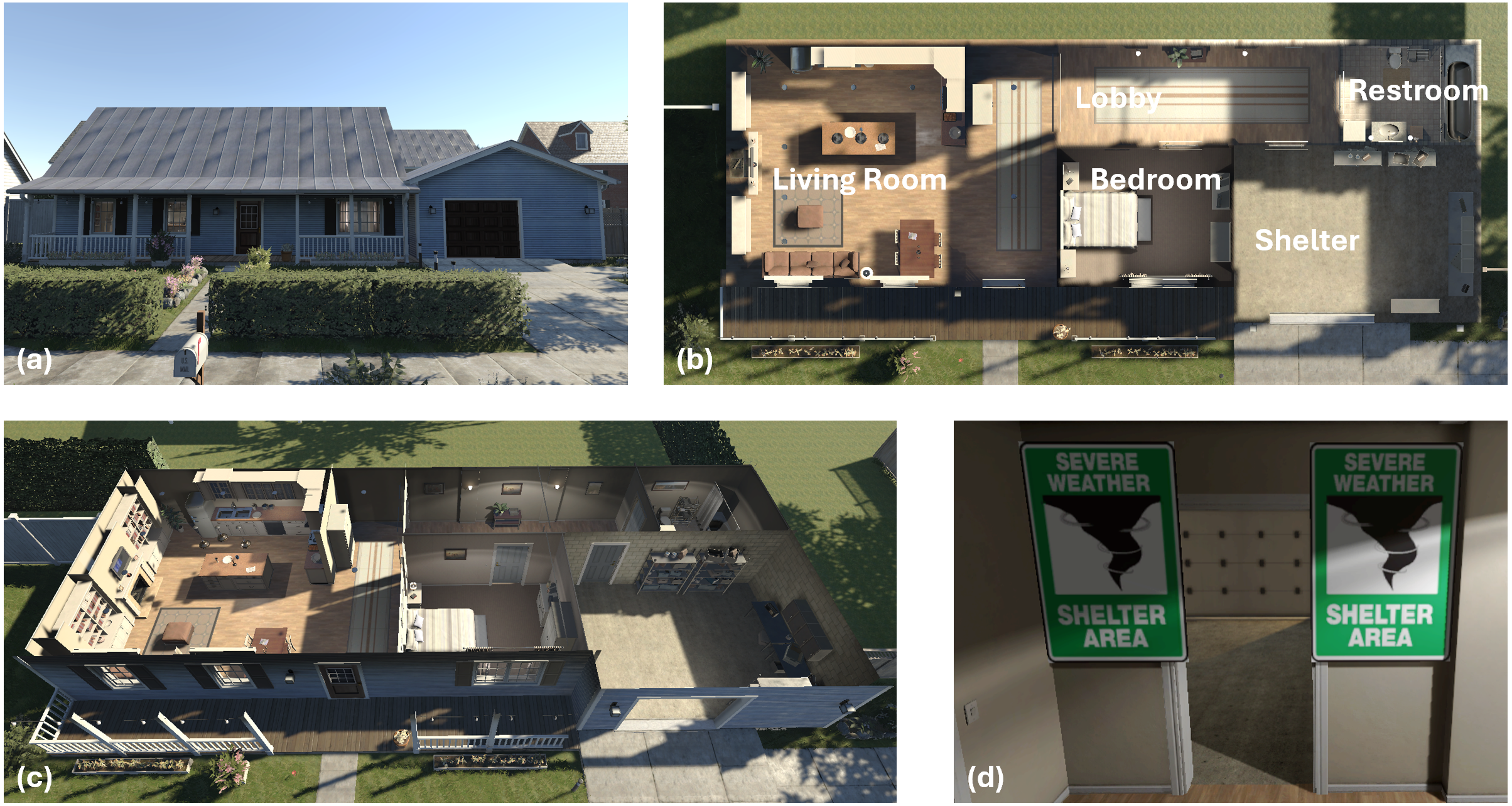}
  \caption{VR environment design: (a) Exterior view of the house. (b) Scene distribution. (c) Interior view of the house. (d) Shelter sign.}
  \label{fig:vr_env}
  \Description{}
\end{figure}

In our system, the VR scene is set in a peaceful neighborhood where the participant engages within the living room of their own virtual home. The virtual house is a single-story structure without stairs, designed to simplify the environment and reduce the risk of dizziness for first-time VR users. The house includes a bathroom, a bedroom, and a garage, with the garage's corner designated as a "safe zone" within the VR scenario. This safe zone represents an area where the participant would be protected from harm in the event of a tornado. The participant must traverse a hallway from the living room to reach the safe zone. The garage door is marked with two "shelter area" signs, indicating the shelter's location. The design of the house's layout follows the Federal Emergency Management Agency (FEMA)\cite{Ready}.. Figure \ref{fig:vr_env} presents the VR environment in detail.

\begin{figure}[h]
  \centering
  \includegraphics[width=\linewidth]{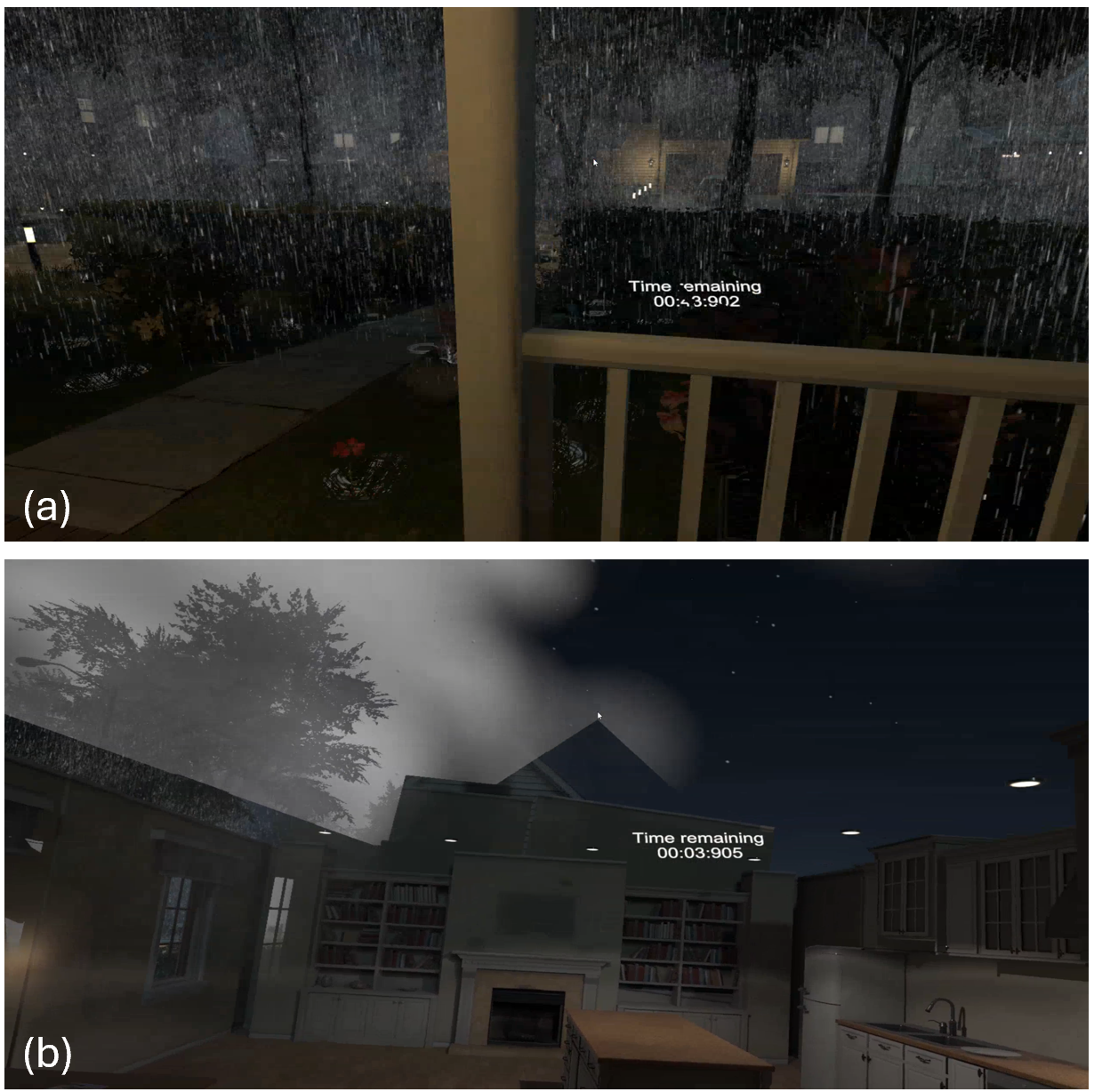}
  \caption{Immersive effect design: (a) Rain. (b) Tornado.}
  \label{fig:imm_env}
  \Description{}
\end{figure}

For a more immersive experience, the VR scene was designed to mirror the actual weather conditions typically associated with a tornado. The scenario was set on a stormy night, along with constant rain, lightning, and thunder, to evoke increased anxiety and fear in participants, simulating the onset of an actual tornado. The thunder sounds were randomized, ensuring that there were no predictable patterns, which further enhanced the realism and unpredictability of the experience. Participants are not confined to the house and are encouraged to explore the environment freely. A thunderous sound is played when a tornado warning is issued, accompanied by haptic feedback as a reminder. Figure \ref{fig:imm_env} shows some examples of immersive design. If the participants stay out of the shelter when the tornado appears, they will witness the destructive effects of the tornado on their virtual home, such as the roof being torn off and furniture being blown away. On the contrary, if the participant is within the safe zone, they will only perceive the loud sound of the tornado, without witnessing any damage. 

\subsubsection{Modality Design in VR}
\begin{figure*}[h]
  \centering
  \includegraphics[width=\linewidth]{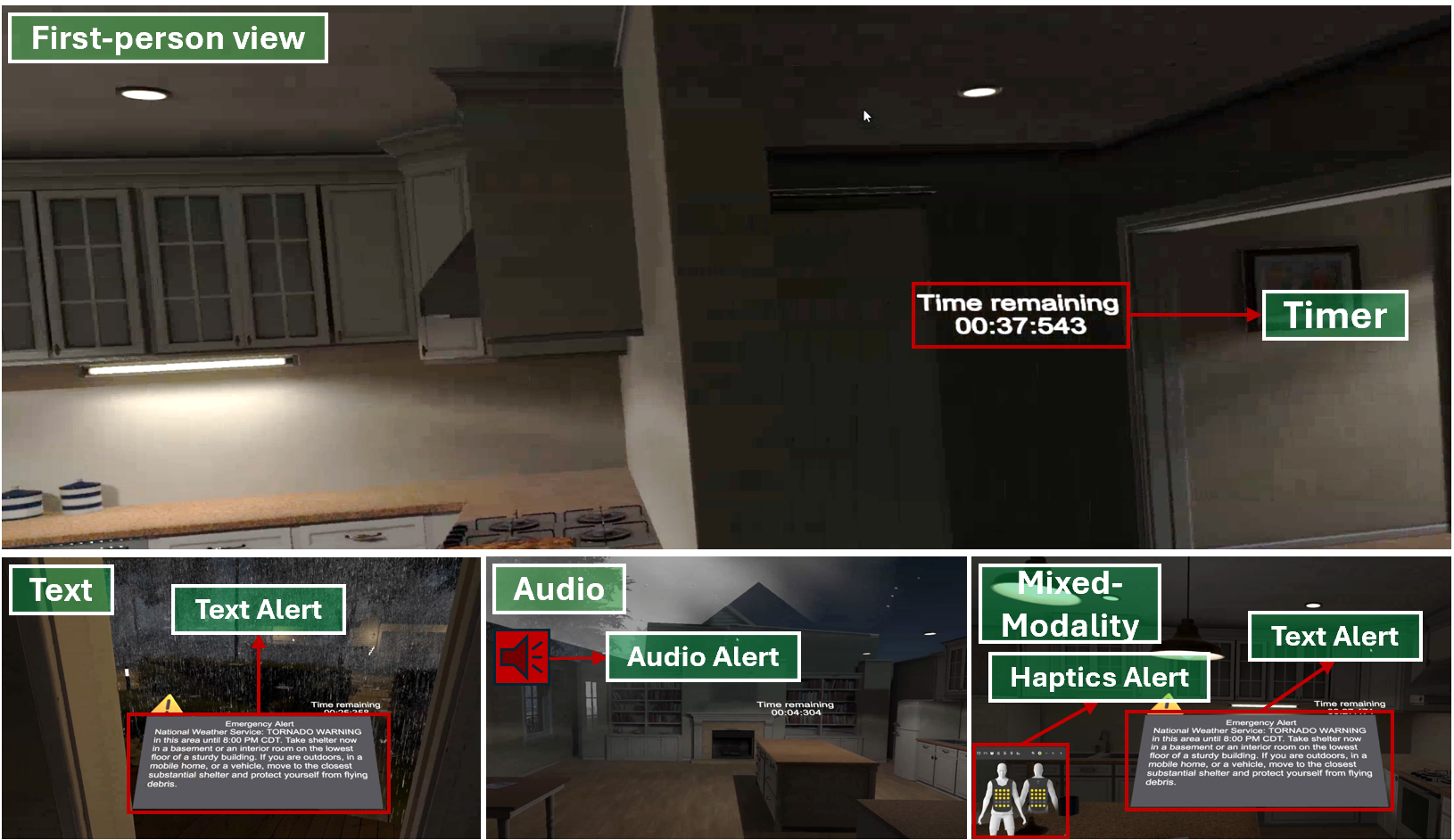}
  \caption{Modality design in the VR system.}
  \label{fig:modality}
  \Description{}
\end{figure*}

In our VR system, we employ three modalities to deliver risk communication alerts: text, audio, and a combination of haptic feedback and text and across all three modalities, a one-minute countdown timer will be consistently present, as shown in Figure \ref{fig:modality}. The countdown timer, incorporated into the study as part of the VR experiment, emphasizes the urgency of the experiment and supports participants in their decision-making process. The design of the content in these three modalities reflects the multi-modality application, which aims to enhance the sense of presence for participants within the virtual environment.

\textbf{Text message.}
The text treatment in this study mimics the traditional method of tornado warning emergency alerts typically received by residents in affected areas on their cell phones during a tornado event in the U.S. The text-based alert displays messages within a box that moves in sync with the user's head movements, ensuring that the alert remains within their field of vision, similar to real-life phone notifications. This warning message is delivered with high certainty that a tornado is approaching, following a series of tornado watch messages informing residents of the likelihood of a tornado occurring that day. In the virtual environment of this experiment, the warning message appears directly in the participant's view. This message box appears for 20 seconds, then disappears for 10 seconds before reappearing, ensuring that it does not obstruct the user's view for an extended period and that the participants remain aware of the ongoing threat.

\textbf{Audio broadcast.}
The audio treatment is delivered as a broadcast through the VR headset, featuring the content of the National Weather Service alert message accompanied by a beeping alarm. This treatment is designed to mimic emergency broadcasts and warning signals, capturing the user's attention through intermittent repetitions throughout the virtual environment. The content of the broadcast is the same as the text message. The broadcast for the alert message will last for 20 seconds, and then it will keep playing the NOAA siren until the trial ends. This method ensures that users are continually aware of potential risks. 

\textbf{Mixed Modality.}
The mixed modality combines the text and haptics modality. The text treatment is the same. The haptics sensation in VR serves as an enhancement that simulates the sense of touch, thereby improving the interaction between participants and the virtual environment. In this study, the haptic feedback is transmitted through the TactSuit X40, a vest worn by participants, which provides haptics sensations corresponding to environmental sounds, such as raindrops and thunder, particularly during the simulated stormy night scenario. This combination of haptics and other sensations aims to create a more immersive and responsive alert system, improving the user's overall situational awareness in the virtual environment.

\subsection{Experiment design}
\begin{figure*}[h]
  \centering
  \includegraphics[width=\linewidth]{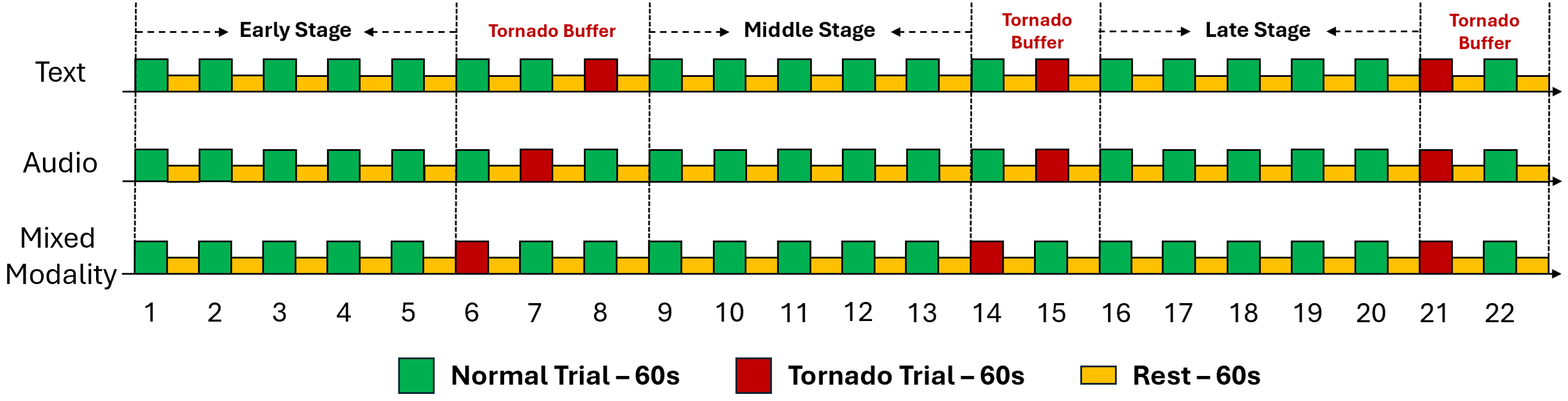}
  \caption{Experiment design. The order of the treatments and the trial in which the tornado occurs are random.}
  \label{fig:exp_design}
  \Description{}
\end{figure*}

This study was designed as a 3 (modality type) $\times$ 3 (trail stage) $\times$  6 (modality sequence) mixed design. The \textit{modality sequence} was designed as a between-subject variable to prevent carryover effects among different modality types. During the experiment, participants will be randomly assigned to one of the six modality sequences (\uppercase\expandafter{\romannumeral1} to \uppercase\expandafter{\romannumeral6}), as described in Table~\ref{tab:group}.  \textit{Modality type} and \textit{trial stage} were designed as the within-subject factors. The types of modality in VR include text, audio, and mix, as introduced in Section 3.1. We designed 22 trials for each type of modality to test the individuals' mitigation behavior and the effect of risk habitation. Since participants might predict the tornado trials for each modality treatment if the tornado trails are fixed, we designed different trial stages and tornado buffers in this experiment. We divided the 22 trials into three trial stages, which are the early stage - before the first tornado trial (trial 1 to trial 5), the middle stage - between the first tornado trial and second tornado trial (trial 9 to trial 13), and late stage between the second tornado trial and third tornado trial (trial 16 to trial 20). In addition, we designed three tornado buffers(trial 6 to trial 8, trial 14 to trial 15, and trial 21 to trial 22) across different modality treatments to prevent participants from predicting tornado trials in each modality treatment. Figure \ref{fig:exp_design} illustrates the details of the experimental design. 

\begin{table}
  \caption{Experiment group.}
  \label{tab:group}
  \begin{tabular}{cc}
    \toprule
    No. & Treatment Sequence \\
    \midrule
    \uppercase\expandafter{\romannumeral1} & text - audio - mixed (T-A-M)\\
    \uppercase\expandafter{\romannumeral2} & text - mixed - audio (T-M-A)\\
    \uppercase\expandafter{\romannumeral3} & audio - text - mixed (A-T-M)\\
    \uppercase\expandafter{\romannumeral4} & audio - mixed - text (A-M-T)\\
    \uppercase\expandafter{\romannumeral5} & mixed - text - audio (M-T-A)\\
    \uppercase\expandafter{\romannumeral6} & mixed - audio - text (M-A-T)\\
  \bottomrule
\end{tabular}
\end{table}

\subsection{Experimental Procedure}
\begin{figure*}[h]
  \centering
  \includegraphics[width=\linewidth]{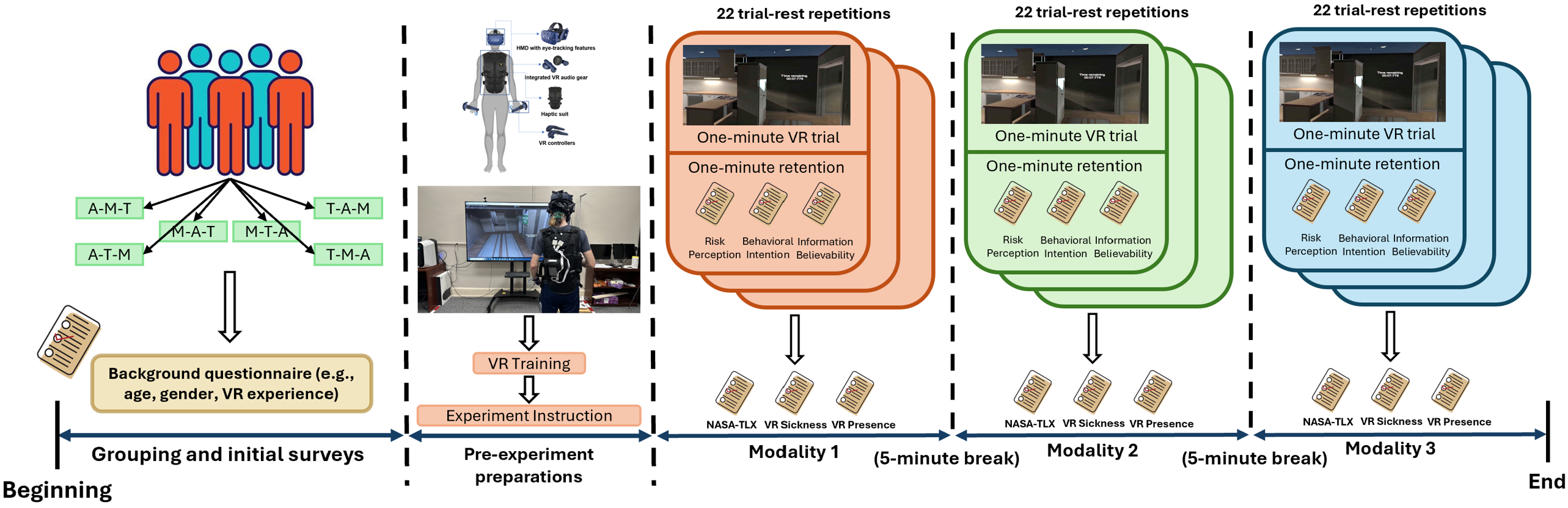}
  \caption{Experimental procedure.}
  \label{fig:expd}
  \Description{}
\end{figure*}

According to our experiment design, the entire experiment will last approximately 4 hours for each participant. After signing informed consent, participants were randomly assigned to one of six groups with different sequences of modalities. Before the experiment sessions, we collected background information from participants that might influence task performance, such as demographic information, disaster experience, gaming experience, and VR experience. We also conducted a training session for participants to familiarize themselves with VR devices, eye tracking systems, TactSuit X40 haptic suit, locomotion, and interaction methods in the virtual environment. During the VR experiment session, each modality treatment consists of 22 trials with a one-minute treatment period and a one-minute rest interval. During rest intervals, participants were asked three questions related to risk perception, behavioral intention, and trust of risk communication messages they received. There was a 5-minute break between each modality treatment. Participants were informed to complete the simulator sickness questionnaire (SSQ)\cite{kennedy1993simulator,balk2013simulator}, VR presence questionnaire\cite{witmer1998measuring,schwind2019using}, and NASA Task Load Index (NASA-TLX)\cite{hart1988development,hart2006nasa} questionnaire, for evaluating their VR sickness, VR presence, and mental workload after each modality treatment. Figure \ref{fig:expd} illustrates the details of the experiment procedure. 

\subsection{Study Measures}
\subsubsection{Shelter Performance}
During each one-minute trial, participants will receive a tornado alert that prompts them to seek shelter within the designated shelter area of their virtual house. The shelter performance is measured by two metrics: the aggregated shelter time and the first shelter entry time. The aggregated shelter time represents the cumulative time that the participant stays in the shelter during each trial. The first shelter entry time represents the time when the participant first enters the shelter after receiving the warning messages. In our VR system, a second-hand timer is constantly displayed within the participants' field of view. Therefore, the recording of the aggregated time and the first shelter entry time should be very simple and accurate. It is worth noting that participants can decide for themselves when to enter or exit the shelter. Hence, during a single trial, participants may enter and exit the shelter multiple times. The aggregated shelter time should record the cumulative time spent in the shelter, whether it occurs once or multiple times. 

\subsubsection{Cognitive Load Index}
We measured the cognitive load with pupil size. According to recent neurofunctional studies, the underlying principle is that activities in the neuromodulatory brain systems can trigger sympathetic and parasympathetic branches of the nerve system, and further trigger sphincter and dilator muscles that control the pupillary changes\cite{van2018pupil,verguts2015adaptive}. The association between pupillary size and cognitive load has been examined in various domains, such as solving mathematics\cite{gavas2017estimation,peysakhovich2017impact}, driving behaviors\cite{palinko2010estimating}, and speech perception\cite{zekveld2011cognitive}. In this study, Tobii eye tracker\cite{TP} was integrated with an HTC VIVE headset to collect pupillary change at a frequency of 120 Hz with an estimated accuracy of 0.5°, which is commonly used for cognitive load estimates\cite{ye2022cognitive, gavas2017estimation, krejtz2018eye}. The general processing pipeline starts with outlier removal, interpolation, and smoothing, then calibrates with brightness effects. Light intensity is the dominant factor for pupil size changes while cognitive changes impose relatively minor impacts\cite{van2018pupil}. To rule out the pupil size change induced by brightness variations, we performed a 1-min pupil size baseline measurement session for each participant in which participants were sitting calmly with varying light intensity. Pierre et al.\cite{pierre2015luminance} find that the brightness received by human eyes is correlated with the relative strength of the RGB values. As such, the real-time brightness is calculated following previous studies \cite{zhou2023cognition, pierre2015luminance}. Pupillary change induced by brightness can be approximated by Equation \ref{pbr}\cite{blackie1999extension}:  
\begin{equation}
P_{Br} = a{e^{b*{Br}}}+c
\label{pbr}
\end{equation}
where $a$, $b$, $c$ are three constants and $Br$ is the calculated brightness.
The light effects were filtered out by subtracting $P_{Br}$ and baseline pupil size from the measured pupil sizes. The resulting standardized values, Cognitive Load Index (CLI), indicate the pupil changes that were purely driven by cognitive load \cite{shi2020impact,xu2024augmented}.

\subsubsection{The evaluation of the VR system}
After each treatment, participants will be requested to complete the Simulator sickness questionnaire (SSQ) \cite{kennedy1993simulator}, the VR presence questionnaire\cite{witmer1998measuring}, and the NASA-TLX questionnaire\cite{hart2006nasa} to evaluate our VR system. The SSQ is used to assess the symptoms of sickness that participants might experience during or after using VR. VR Presence Questionnaire is used to measure a user's sense of "presence" or "immersion" in a virtual reality environment. NASA-TLX is used to measure perceived workload in human performance contexts, including virtual reality environments. It evaluates the cognitive and physical demands placed on users while completing a task. 

\subsubsection{Subjective Questions}
After each experiment trial, the participants will be asked three questions to measure their risk perception, proactive response, and trust in the alert. The responses to these questions will be recorded, using a Likert Scale ranging from 1 (not at all likely) to 10 (extremely likely). The questions were as follows:

\begin{itemize}
\item How possible can a tornado happen in the subsequent trial? (Risk perception)
\item How likely are you to seek shelter in the subsequent trial? (Behavioral intention)
\item How much do you trust the risk communication you received in this trial? (Information believability)
\end{itemize}

\section{RESULTS}

\subsection{Participants}
We recruited 27 participants from the University of Alabama (21 males and 6 females). Among the participants, three participants (1 male and 2 female) were unable to finish the experiment due to VR sickness during the experiment, and their incomplete data were removed from the data analysis. The ages of the participants ranged from 18 to 44 years, with a mean age (M) of 29.47 years and a standard deviation (SD) of 4.73. Participants' prior experiences with natural hazards and gaming/VR were also collected using a 3-point Likert scale (where 1 = no experience, 2 = normal experience, and 3 = extensive experience). The results indicated that the participants have less than normal experience of natural hazards and also do not have related gaming or VR experience.

% \begin{table*}
%   \caption{Prior Experience of Participants}
%   \label{tab:participants}
%   \begin{tabular}{cccc}
%     \toprule
%     N=24 & Types & Mean & Standard Deviation \\
%     \midrule
%     \multirow{5}*{Disaster experience} & Tornado & 1.50 & 0.50\\
%     ~ & Hurricane & 1.21 & 0.41\\
%     ~ & Flooding & 1.58 & 0.57\\
%     ~ & Earthquake & 1.54 & 0.64\\
%     ~ & Wildfire & 1.08 & 0.27\\
%     \midrule
%     \multirow{4}*{Gaming experience} & Computer or video & 2.17 & 0.47\\
%     ~ & Virtual Reality & 1.63 & 0.56\\
%     ~ & First person-shooter & 1.75 & 0.52\\
%     ~ & Third-person role-playing & 1.75 & 0.60\\
%   \bottomrule
% \end{tabular}
% \end{table*}

\subsection{Shelter Performance}
Our experiment follows a three-way mixed experiment design with a between-subject factor (sequence) and two within-subject factors (modality and stage). We used the \textit{ggplot2} package\cite{ggplot2} in the R to illustrate the experiment results.

\subsubsection{Aggregated Shelter Time}
\begin{figure}[h]
  \centering
  \includegraphics[width=\linewidth]{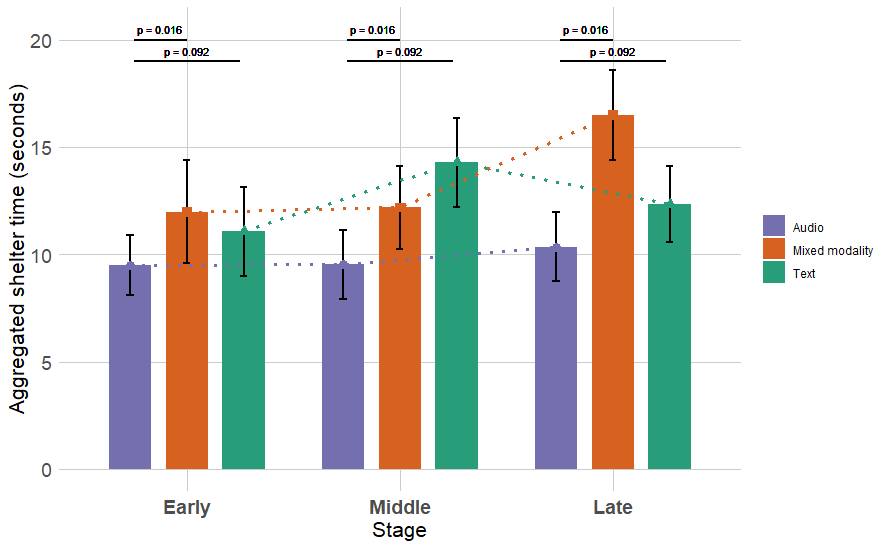}
  \caption{Task performance: Aggregated shelter time.}
  \label{fig:aggregated}
  \Description{}
\end{figure}

Our data met the normality assumption using the Shapiro-Wilk test and the homogeneity assumption using Levene's test of Analysis of variance (ANOVA). As a result, we used a three-way mixed ANOVA to compare the differences in aggregated shelter time with a significant difference ($\alpha$) of 0.05. We found significant differences in one within-subject factor (modality) of aggregated shelter time ($F(2,36) = 4.623, p = 0.016, \eta^2 = 0.041$) as illustrated in Figure \ref{fig:aggregated} and an interaction effect (sequence:modality) ($F(10,36) = 3.768, p = 0.002, \eta^2 = 0.147$). There was no between-subject effect (sequence) ($p = 0.337$), no within-subject effect (stage) ($p = 0.261$) or other interaction effects. Based on the post-hoc paired Tukey's Honest Significant Difference (HSD) method, we found a significant difference between the audio modality and the haptics modality ($p = 0.016$). There was a marginal difference in the aggregated shelter time between the audio modality and the text modality ($p = 0.092$). However, there is no difference between the haptics modality and the text modality ($p = 0.723$). To explore the differences in aggregated shelter time at different modalities for each stage, we also performed a two-way mixed ANOVA with a significant difference ($\alpha$) of 0.05. For the early stage, there is no significant difference. However, we found significant differences in within-subject factor (modality) for both the middle stage ($F(2,36) = 3.620, p = 0.037, \eta^2 = 0.065$) and the late stage ($F(2,36) = 3.950, p = 0.028, \eta^2 = 0.104$). Based on the post-hoc paired Tukey's Honest Significant Difference (HSD) method, we found a significant difference between the audio modality and the text modality ($p = 0.029$) for the middle stage and a significant difference between the audio modality and the mixed modality ($p = 0.024$) for the late stage.

\subsubsection{First Shelter Entry Time}
\begin{figure}[h]
  \centering
  \includegraphics[width=\linewidth]{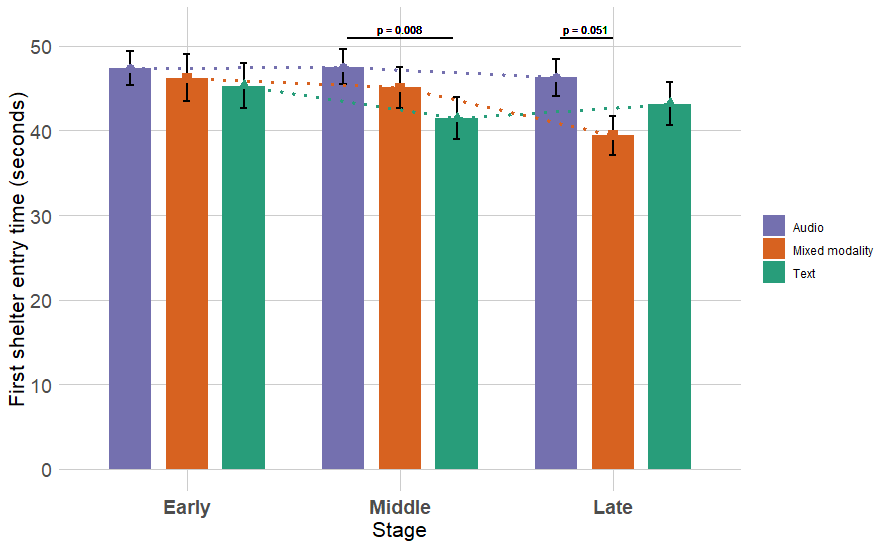}
  \caption{Task performance: The first time that the participant enters the shelter after the alert is delivered.}
  \label{fig:first}
  \Description{}
\end{figure}

Our data met the normality assumption using the Shapiro-Wilk test and the homogeneity assumption using Levene's test of ANOVA. As a result, we used a three-way mixed ANOVA to compare the differences in the first shelter entry time with a significant difference ($\alpha$) of 0.05. We found a significant difference in one within-subject factor (modality) of the first shelter entry time ($F(2,36) = 3.481, p = 0.041, \eta^2 = 0.029$) as illustrated in Figure \ref{fig:first} and an interaction effect (sequence:modality) ($F(10,36) = 2.901, p = 0.009, \eta^2 = 0.110$). There was no between-subject effect (sequence) ($p = 0.353$), no within-subject effect (stage) ($p = 0.102$) or other interaction effects. Based on the post-hoc paired Tukey's Honest Significant Difference (HSD) method, we did not find any significant differences between different modalities. There were marginal differences in the aggregated shelter time between the audio modality and the text modality ($p = 0.059$) and between the audio modality and haptics modality ($p = 0.086$). However, there was no difference between the haptics modality and the text modality ($p = 0.984$). Based on the
post-hoc paired t-test (Bonferroni correction), we found significant differences between the audio modality and the haptics modality ($p = 0.041$) and between the text modality and the haptics modality ($p = 0.007$) under sequence = T-A-H and a marginal difference between the audio modality and the text modality ($p = 0.06$) under sequence = A-H-T. To explore the differences in the first shelter entry time at different modalities for each stage, we also performed a two-way mixed ANOVA with a significant difference ($\alpha$) of 0.05. For the early stage, there is no significant difference. However, we found a significant difference in within-subject factor (modality) for the middle stage ($F(2,36) = 5.165, p = 0.011, \eta^2 = 0.066$). We also found a marginal difference in within-subject factor (modality) for the late stage ($F(2,36) = 2.982, p = 0.063, \eta^2 = 0.075$). Based on the post-hoc paired Tukey's Honest Significant Difference (HSD) method, we found a significant difference between the audio modality and the text modality ($p = 0.008$) for the middle stage and a marginal difference between the audio modality and the mixed modality ($p = 0.051$) for the late stage.

\subsection{Risk Perception}
Our data did not meet the normality assumption of ANOVA using the Shapiro-Wilk test. As a result, we used ART (Aligned Rank Transform) which is a non-parametric data analysis technique to compare the differences in the level of risk perception with a significant difference ($\alpha$) of 0.05. We found a significant difference in one interaction effect (sequence:modality) ($F(10,144) = 2.150, p = 0.024$). There were no main effects or other interaction effects. Based on the post-hoc paired t-test (Bonferroni correction), we only found a marginal difference between the text modality and the haptics modality ($p = 0.059$) under sequence = T-A-H.

% \begin{figure}[h]
%   \centering
%   \includegraphics[width=\linewidth]{Risk_Perception.png}
%   \caption{Risk perception: How possible is it that a tornado will happen in the next trial? (A) The result for Sequence = T-A-H. (B) The overall result.}
%   \label{fig:risk}
%   \Description{}
% \end{figure}

\subsection{Behavioral Intention}
\begin{figure}[h]
  \centering
  \includegraphics[width=\linewidth]{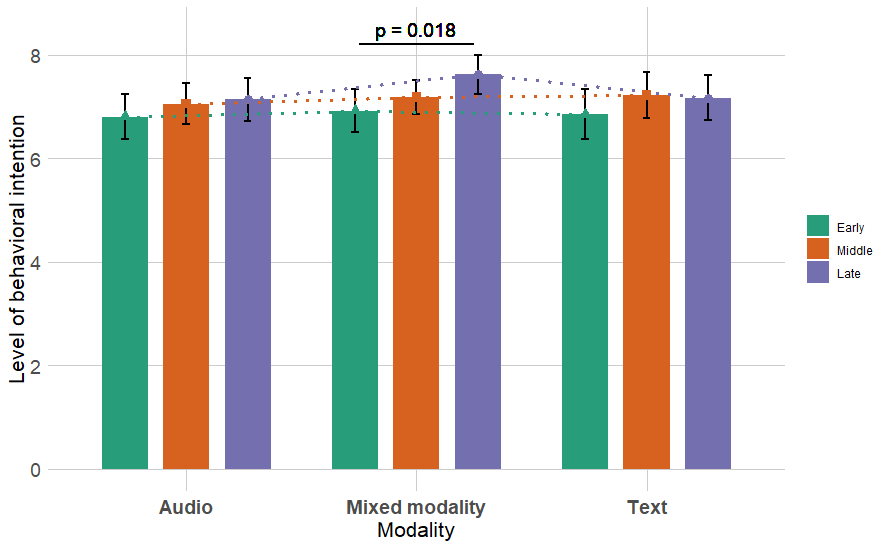}
  \caption{Behavioral intention: At this point in time, how likely are you to seek shelter?}
  \label{fig:behav}
  \Description{}
\end{figure}

Our data met the normality assumption using the Shapiro-Wilk test and the homogeneity assumption using Levene's test of ANOVA. As a result, we used a three-way mixed ANOVA to compare the differences in the first shelter entry time with a significant difference ($\alpha$) of 0.05. We found a significant difference in one within-subject factor (stage) of the level of behavioral intention ($F(2,36) = 4.654, p = 0.016, \eta^2 = 0.011$) as illustrated in Figure \ref{fig:behav}. There was no between-subject effect (sequence) ($p = 0.414$), no within-subject effect (modality) ($p = 0.704$) or other interaction effects. Based on the post-hoc paired Tukey's Honest Significant Difference (HSD) method, we found a significant difference between the early stage and the late stage ($p = 0.013$). However, there was no difference between the early stage and the middle stage ($p = 0.136$) or between the middle stage and the late stage ($p = 0.556$). To explore the differences in the behavioral intention at different stages for each modality, we also performed a two-way mixed ANOVA with a significant difference ($\alpha$) of 0.05. For the audio modality and the text modality, there is no significant difference. However, we found a significant difference in within-subject factor (stage) for the mixed modality ($F(2,36) = 4.208, p = 0.023, \eta^2 = 0.032$). Based on the post-hoc paired Tukey's Honest Significant Difference (HSD) method, we found a significant difference between the early stage and the late stage ($p = 0.018$) the mixed modality.

\subsection{Information Believability}
Our data did not meet the normality assumption of ANOVA using the Shapiro-Wilk test. As a result, we used ART (Aligned Rank Transform) which is a non-parametric data analysis technique to compare the differences in the level of information believability with a significant difference ($\alpha$) of 0.05. We found a significant difference in one interaction effect (sequence:modality) ($F(10,144) = 2.150, p = 0.000$). There were no main effects or other interaction effects. Based on the post-hoc paired t-test (Bonferroni correction), we did not find any difference between the modalities under different sequences.

% \begin{figure}[h]
%   \centering
%   \includegraphics[width=\linewidth]{Information.png}
%   \caption{Information believability: How much do you trust the risk communication information you have seen?}
%   \label{fig:info}
%   \Description{}
% \end{figure}

\subsection{Eye Tracking}
\begin{figure}[h]
  \centering
  \includegraphics[width=\linewidth]{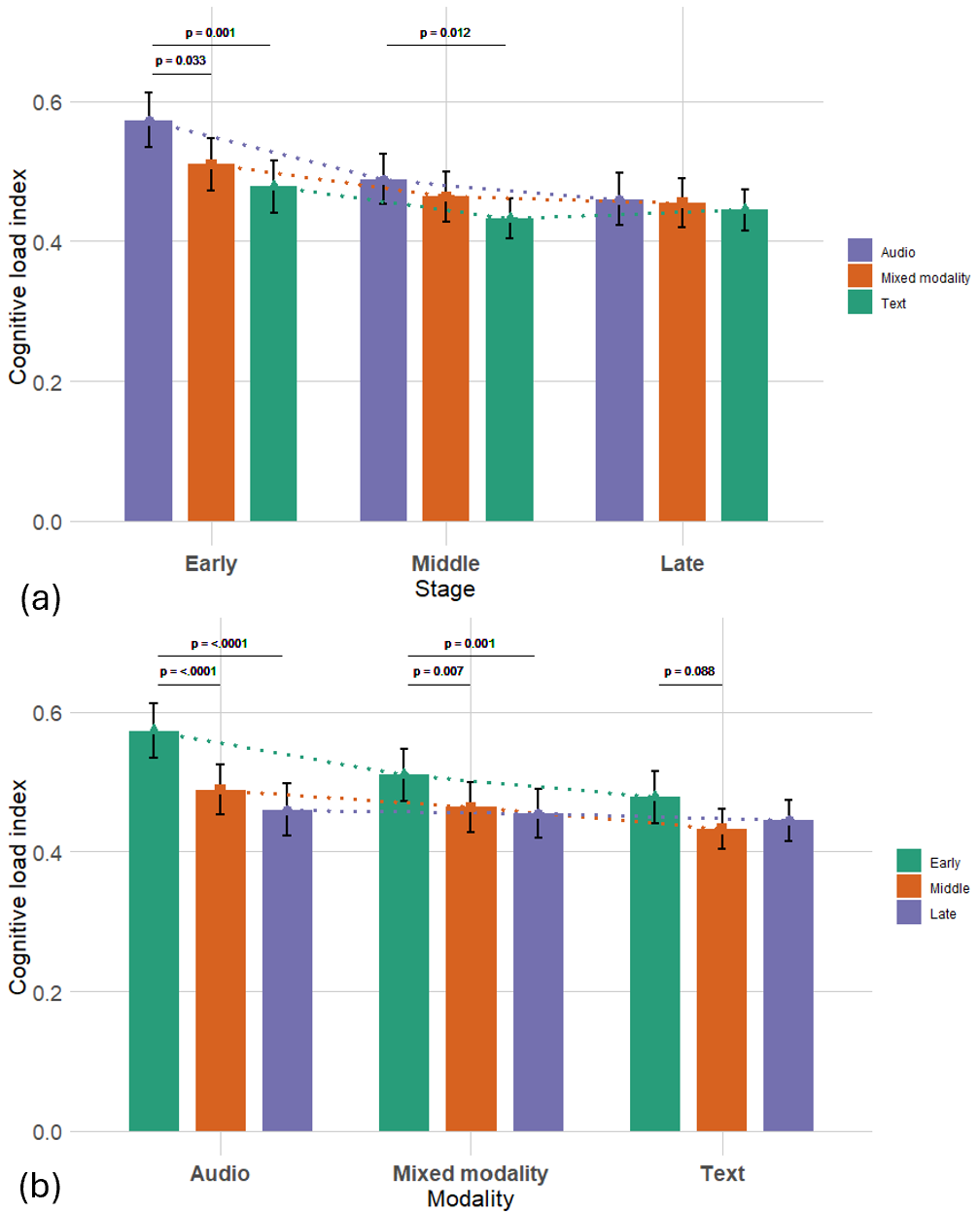}
  \caption{Eye tracking: (a) Modality. (b) Stage.}
  \label{fig:eye}
  \Description{}
\end{figure}

Our data met the normality assumption using the Shapiro-Wilk test and the homogeneity assumption using Levene's test of Analysis of variance (ANOVA). As a result, we used a three-way mixed ANOVA to compare the differences in the cognitive load index with a significant difference ($\alpha$) of 0.05. We found significant differences in both within-subject factor (modality) ($F(2,36) = 6.773, p = 0.003, \eta^2 = 0.022$) and within-subject factor (stage) of the cognitive load index ($F(2,36) = 33.068, p = 0.000, \eta^2 = 0.037$) as illustrated in Figure \ref{fig:eye}. We also found an interaction effect (sequence:modality) ($F(10,36) = 2.721, p = 0.013, \eta^2 = 0.043$) and an interaction effect (modality:stage) ($F(4,72) = 3.363, p = 0.014, \eta^2 = 0.008$) as illustrated in Figure \ref{fig:aggregated}. There was no between-subject effect (sequence) ($p = 0.618$) or other interaction effects. Based on the post-hoc paired Tukey's Honest Significant Difference (HSD) method, we found a significant difference between the audio modality and the text modality ($p = 0.002$). However, we did not find any other difference. Based on the post-hoc paired t-test (Bonferroni correction) in sequence:modality, we found significant differences between the audio modality and the text modality under sequence = A-H-T ($p = 0.044$) and sequence = A-T-H ($p = 0.048$) and a marginal difference between audio modality and the haptics modality under sequence = A-H-T ($p = 0.097$). Based on the post-hoc paired t-test (Bonferroni correction) in modality:stage, we did not find any difference between the stages under different modalities.

\subsection{VR System Evaluation}
\begin{figure}[h]
  \centering
  \includegraphics[width=\linewidth]{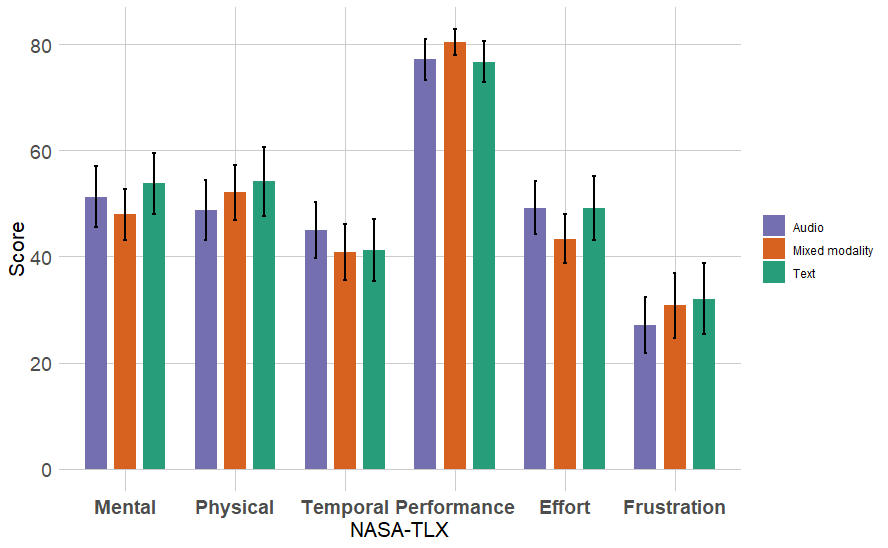}
  \caption{The result of NASA-TLX questionnaire.}
  \label{fig:participants}
  \Description{}
\end{figure}

Our NASA-TLX data met the normality assumption using the Shapiro-Wilk test and the homogeneity assumption using Levene's test of ANOVA. The two-way mixed ANOVA did not find any main effect or interaction effect but we still visualized the result in Figure \ref{fig:participants}. The results of VR sickness show that the audio modality induces the most severe nausea among the three modalities, while the text modality and mixed modality result in comparable levels of nausea. Both the audio modality and text modality cause more severe oculomotor symptoms compared to the mixed modality. Furthermore, the audio modality leads to the most severe disorientation, followed by the mixed modality, with the text modality causing the least disorientation. By calculating the total sickness score, we found that the audio modality resulted in the highest sickness score (626.14), while the text modality and mixed modality resulted in lower sickness scores (547.12 and 547.76, respectively). The specific details of VR sickness are presented in Table~\ref{tab:sickness}. The results of the VR presence show no significant differences among the three different modalities. The total VR presence scores for the audio, text, and mixed modalities were 115.75, 114.46, and 116.54, respectively. The results of VR presence are presented in Table \ref{tab:presence} in detail.

% \begin{table}
%   \caption{VR Sickness Summary (M, Standard Deviation)}
%   \label{tab:sickness}
%   \begin{tabular}{cccc}
%     \toprule
%      & Audio & Haptics & Text \\
%     \midrule
%     General discomfort & (1.33, 0.92) & (1.21, 0.78) & (1.50, 0.88) \\
%     Fatigue & (1.21, 0.88) & (1.08, 0.79) & (1.50, 1.10) \\
%     Headache & (0.88, 0.85) & (0.79, 0.88) & (0.92, 0.97) \\
%     Eye strain & (1.04, 0.91) & (0.96, 0.95) & (0.96, 0.91) \\
%     Difficulty focusing & (0.71, 0.86) & (0.71, 1.00) & (0.67, 0.70) \\
%     Increased salivation & (0.50, 0.88) & (0.58, 0.88) & (0.42, 0.78) \\
%     Sweating & (0.46, 0.83) & (0.38, 0.82) & (0.38, 0.71) \\
%     Nausea & (0.92, 1.25) & (0.75, 1.03) & (0.75, 1.03) \\
%     Difficulty concentrating & (0.71, 0.75) & (0.54, 0.83) & (0.50, 0.66) \\
%     Fullness of head & (0.92, 0.93) & (0.96, 0.95) & (0.96, 0.95) \\
%     Blurred vision & (0.71, 0.86) & (0.67, 0.82) & (0.46, 0.66) \\
%     Dizzy (eyes open) & (1.04, 1.12) & (0.92, 1.02) & (0.88, 0.95) \\
%     Dizzy (eyes closed) & (0.83, 0.96) & (0.75, 0.94) & (0.79, 0.98) \\
%     Vertigo & (0.71, 0.91) & (0.58, 0.88) & (0.54, 0.83) \\
%     Stomach awareness & (0.75, 0.99) & (0.58, 0.83) & (0.58, 0.93) \\
%     Burping & (0.42, 0.65) & (0.38, 0.88) & (0.21, 0.41) \\
%   \bottomrule
% \end{tabular}
% \end{table}

\begin{table}
  \caption{VR sickness score for each modality.}
  \label{tab:sickness}
  \begin{tabular}{cccc}
    \toprule
     & Audio & Mixed Modality & Text \\
    \midrule
    General discomfort & 1.33 & 1.21 & 1.50 \\
    Fatigue & 1.21 & 1.08 & 1.50 \\
    Headache & 0.88 & 0.79 & 0.92 \\
    Eye strain & 1.04 & 0.96 & 0.96 \\
    Difficulty focusing & 0.71 & 0.71 & 0.67 \\
    Increased salivation & 0.50 & 0.58 & 0.42 \\
    Sweating & 0.46 & 0.38 & 0.38 \\
    Nausea & 0.92 & 0.75 & 0.75 \\
    Difficulty concentrating & 0.71 & 0.54 & 0.50 \\
    Fullness of head & 0.92 & 96 & 0.96 \\
    Blurred vision & 0.71 & 0.67 & 0.46 \\
    Dizzy (eyes open) & 1.04 & 0.92 & 0.88 \\
    Dizzy (eyes closed) & 0.83 & 0.75 & 0.79 \\
    Vertigo & 0.71 & 0.58 & 0.54 \\
    Stomach awareness & 0.75 & 0.58 & 0.58 \\
    Burping & 0.42 & 0.38 & 0.21 \\
    \midrule
    Nausea & 48.50 & 42.14 & 41.34 \\
    Oculomotor & 49.90 & 45.16 & 49.27 \\
    Disorientation & 69.02 & 59.16 & 55.68 \\
    \midrule
    TotalSSQScore & 626.14 & 547.76 & 547.12 \\
  \bottomrule
\end{tabular}
\end{table}

\begin{table}
  \caption{VR presence score for each modality.}
  \label{tab:presence}
  \begin{tabular}{cccc}
    \toprule
     & Audio & Mixed Modality & Text \\
    \midrule
    Realism & 36.46 & 36.21 & 35.38 \\
    Possibility to act & 21.50 & 22.29 & 21.83 \\
    Quality of interface & 13.46 & 14.21 & 13.83 \\
    Possibility to examine & 16.25 & 16.29 & 15.58 \\
    Self-evaluation of performance & 11.75 & 11.29 & 11.38 \\
    Sounds* & 16.33 & 16.25 & 16.46 \\
    \midrule
    TotalPScore & 115.75 & 116.54 & 114.46 \\
  \bottomrule
\end{tabular}
\end{table}

\section{DISCUSSION}
Our study addressed two research questions. First, we investigated the impacts of different modalities in VR on individuals' mitigation behavior (\textbf{RQ1}). We found that, compared to the audio modality and text modality, the mixed modality significantly enhances individuals' mitigation behavior in the VR system. Our second research question explored how different modalities affect risk habituation in the VR system (\textbf{RQ2}). We found that the text modality leads to significant risk habituation, while the audio and mixed modalities do not induce risk habituation. In addition, the mixed modality demonstrated a better effect in reducing risk habituation. We provide a more in-depth discussion of the overall impact of different modalities on individuals' mitigation behavior and their habituation to risk in the following sections.

\subsection{The Effect of Different Modalities on Shelter Performance}
The shelter performance was evaluated by two metrics, which are the aggregated shelter time and the first shelter entry time. The results indicated that different modalities result in varying shelter performance. Although there were no significant differences among the three modalities in the early stage of receiving the warning message, differences began to emerge as the trial progressed. In the middle stage, the text modality demonstrated better shelter performance, but was affected by the risk habituation effect. In contrast, the audio modality maintained consistent shelter performance throughout all stages. However, the mixed modality showed resilience to the risk habituation effect, with the shelter performance of the participants improving over time. In the later stage, the mixed modality significantly outperformed both the audio and text modalities. These findings suggested that the mixed modality can improve individuals' shelter behavior and reduce the impact of risk habituation in the context of natural hazards risk communications. The better performance of the "mixed modality" in our experiment could be primarily due to two reasons: (1)multimodal systems integrate different forms of communication, which helps in improving situational awareness, and ensuring that critical information is better understood, especially during high-stress scenarios\cite{algiriyage2022multi}, (2)using multiple sensory inputs (e.g., audio, text, and haptic) reduces the likelihood of individuals becoming desensitized to warnings, maintaining their attention and responsiveness over time\cite{chiozzi2023natural}.

\subsection{The Effect of Risk Perception, Behavioral Intention and Information Believability on Shelter Performance}
We tried to interpret shelter performance through participants' risk perception, proactive behavior, and trust in information. Our results showed that participants' risk perception and information believability did not have significant differences between different stages and modalities, indicating that the effects of different modalities on shelter performance were not influenced by risk perception or information believability. We found a significant difference between the early stage and the late stage in the results of behavioral intention. The results showed that the participants' behavior intention did not change significantly in the audio and text modalities between different stages. However, the mixed modality led to significantly higher behavioral intention in the late stage compared to the early stage, which is consistent with the shelter performance results.

\subsection{Eye-tracking data}
We also collected and analyzed eye-tracking data. The results show that cognitive load differed across modalities, and the same modality exhibited significant variations in cognitive load across different stages. At the early stage, the cognitive load for the audio modality was significantly higher than that for the text and mixed modalities. This is because, unlike the text and mixed modalities, the audio modality lacks textual information to focus on when the tornado alert is present. Participants would unconsciously observe the surrounding environment, leading to a significantly higher cognitive load. However, as the experiment progressed, participants became accustomed to the VR system, resulting in a noticeable reduction in cognitive load across all modalities. The gap between different modalities also began to narrow. At the late stage, the cognitive load across different modalities was almost the same. Since both the mixed modality and text modality provided the same text alert, there was no significant difference in their cognitive load. However, because the mixed modality also delivered haptics alerts, its cognitive load appeared slightly higher than that of the text modality, though the difference was not obvious.

\subsection{VR System Evaluation}
The NASA-TLX results show no significant differences in the perceived workload across different modalities. The VR presence results show no significant differences in the level of presence or immersion in the VR across different modalities. However, the VR sickness results indicated that the audio modality induced a more noticeable sensation of sickness compared to the text modality and mixed modality, especially in nausea and disorientation symptoms. The results of the three questionnaires for the text modality and mixed modality are generally consistent, indicating that adding haptics feedback to the text modality does not bring additional workload or VR sickness. In addition, it also dose not enhance VR presence or immersion.

\section{LIMITATIONS AND FUTURE WORK}
There are some limitations in our work. One limitation is that our sample size is not large enough relative to the complexity of our experiment. Although we can find some useful patterns from the experimental results, the differences in some data are not very pronounced (marginal difference or even no difference in statistics) because of the limited sample size. Second, we also acknowledge that the duration of our experiment (approximately 4 hours) may have influenced the results, as participants' attention and engagement in the experiment are likely to decline as the study progresses. At last, due to the experimental setup requiring an intuitive way to inform participants of the approaching tornado, the haptics design was not well-suited for this purpose. Therefore, instead of testing the haptics modality separately, we used a text-haptics mixed modality which makes it impossible for us to directly compare the performance of the haptics modality with audio or haptics modality with text in the experiment.

Future work could continue exploring the performance of other modalities. At the same time, researchers should collect additional types of data (such as heart rate data, emotional data and fNIRS data) to provide a more comprehensive explanation of the performance of different modalities. Future work should also aim to explore the long-term impacts of natural hazards, where VR integrated with effective modalities could mitigate risk habituation. Understanding these effects and underlying mechanisms may lead to the development of optimized technologies for reducing losses caused by natural hazards. 

\section{CONCLUSION}
In this study, we investigated the impact of various modalities on individuals' mitigation behavior and risk habituation in a VR system. Our results demonstrated that the mixed modality significantly improved mitigation behavior compared to both the audio and text modalities, particularly in the late stage of the experiment. While the text modality led to a clear risk habituation effect, the mixed modality effectively counteracted this phenomenon. Additionally, we found that the mixed modality was indeed more effective in enhancing behavioral proactivity. This is one of the reasons why the mixed modality performed better overall. By integrating eye-tracking technology, we also found that the audio modality, compared to the text modality, increased participants' cognitive load, which implies that their level of vigilance increased. These findings underscore the effectiveness of mixed modality in enhancing risk awareness and response in high-stakes environments such as tornadoes. We hope our work will help to inform future research and practical applications aimed at improving risk mitigation strategies in immersive simulations like VR.

\bibliographystyle{ACM-Reference-Format}
\bibliography{sample-base}


\begin{thebibliography}{88}

%%% the Web address.

\ifx \showCODEN    \undefined \def \showCODEN     #1{\unskip}     \fi
\ifx \showDOI      \undefined \def \showDOI       #1{#1}\fi
\ifx \showISBNx    \undefined \def \showISBNx     #1{\unskip}     \fi
\ifx \showISBNxiii \undefined \def \showISBNxiii  #1{\unskip}     \fi
\ifx \showISSN     \undefined \def \showISSN      #1{\unskip}     \fi
\ifx \showLCCN     \undefined \def \showLCCN      #1{\unskip}     \fi
\ifx \shownote     \undefined \def \shownote      #1{#1}          \fi
\ifx \showarticletitle \undefined \def \showarticletitle #1{#1}   \fi
\ifx \showURL      \undefined \def \showURL       {\relax}        \fi
% invisible to TeX
\providecommand\bibfield[2]{#2}
\providecommand\bibinfo[2]{#2}
\providecommand\natexlab[1]{#1}
\providecommand\showeprint[2][]{arXiv:#2}

\bibitem[HTC({[n.\,d.]})]%
        {HTC}
 \bibinfo{year}{[n.\,d.]}\natexlab{}.
\newblock \bibinfo{title}{HTC VIVE Pro Eye}.
\newblock
\newblock
\urldef\tempurl%
\url{https://www.vive.com/sea/product/vive-pro-eye/overview/}
\showURL{%
\tempurl}


\bibitem[X40({[n.\,d.]})]%
        {X40}
 \bibinfo{year}{[n.\,d.]}\natexlab{}.
\newblock \bibinfo{title}{TactSuit X40}.
\newblock
\newblock
\urldef\tempurl%
\url{https://www.bhaptics.com/tactsuit/tactsuit-x40/}
\showURL{%
\tempurl}


\bibitem[uni({[n.\,d.]})]%
        {unity}
 \bibinfo{year}{[n.\,d.]}\natexlab{}.
\newblock \bibinfo{title}{Unity Engine 2019.4.30f1}.
\newblock
\newblock
\urldef\tempurl%
\url{https://unity.com/releases/editor/whats-new/2019.4.30#notes}
\showURL{%
\tempurl}


\bibitem[NAP(2012)]%
        {NAP13457}
 \bibinfo{year}{2012}\natexlab{}.
\newblock \bibinfo{booktitle}{\emph{Disaster Resilience: A National Imperative}}.
\newblock \bibinfo{publisher}{The National Academies Press}, \bibinfo{address}{Washington, DC}.
\newblock
\showISBNx{978-0-309-26150-0}
\urldef\tempurl%
\url{https://doi.org/10.17226/13457}
\showDOI{\tempurl}


\bibitem[TP(2022)]%
        {TP}
 \bibinfo{year}{2022}\natexlab{}.
\newblock \bibinfo{title}{Tobii Pro VR Integration.}
\newblock
\newblock
\urldef\tempurl%
\url{https://www.tobiipro.com/product-listing/vr-integration/}
\showURL{%
\tempurl}


\bibitem[Rea(2024)]%
        {Ready}
 \bibinfo{year}{2024}\natexlab{}.
\newblock \bibinfo{title}{Ready for Tornadoes}.
\newblock
\newblock
\urldef\tempurl%
\url{https://www.ready.gov/tornadoes}
\showURL{%
\tempurl}


\bibitem[Algiriyage et~al\mbox{.}(2022)]%
        {algiriyage2022multi}
\bibfield{author}{\bibinfo{person}{Nilani Algiriyage}, \bibinfo{person}{Raj Prasanna}, \bibinfo{person}{Kristin Stock}, \bibinfo{person}{Emma~EH Doyle}, {and} \bibinfo{person}{David Johnston}.} \bibinfo{year}{2022}\natexlab{}.
\newblock \showarticletitle{Multi-source multimodal data and deep learning for disaster response: a systematic review}.
\newblock \bibinfo{journal}{\emph{SN Computer Science}}  \bibinfo{volume}{3} (\bibinfo{year}{2022}), \bibinfo{pages}{1--29}.
\newblock


\bibitem[Austin and Jin(2016)]%
        {austin2016social}
\bibfield{author}{\bibinfo{person}{Lucinda Austin} {and} \bibinfo{person}{Yan Jin}.} \bibinfo{year}{2016}\natexlab{}.
\newblock \showarticletitle{Social media and crisis communication: Explicating the social-mediated crisis communication model}.
\newblock In \bibinfo{booktitle}{\emph{Strategic communication}}. \bibinfo{publisher}{Routledge}, \bibinfo{pages}{163--186}.
\newblock


\bibitem[Backfried et~al\mbox{.}(2016)]%
        {backfried2016general}
\bibfield{author}{\bibinfo{person}{Gerhard Backfried}, \bibinfo{person}{Christian Schmidt}, \bibinfo{person}{Dorothea Aniola}, \bibinfo{person}{Christian Meurers}, \bibinfo{person}{Klaus Mak}, \bibinfo{person}{Johannes G{\"o}llner}, \bibinfo{person}{Andreas Peer}, \bibinfo{person}{Gerald Quirchmayr}, \bibinfo{person}{Gerald Czech}, {and} \bibinfo{person}{Markus Glanzer}.} \bibinfo{year}{2016}\natexlab{}.
\newblock \showarticletitle{A general framework for using social and traditional media during natural disasters: Quoima and the central European floods of 2013}.
\newblock In \bibinfo{booktitle}{\emph{Fusion Methodologies in Crisis Management: Higher Level Fusion and Decision Making}}. \bibinfo{publisher}{Springer}, \bibinfo{pages}{469--487}.
\newblock


\bibitem[Bailenson et~al\mbox{.}(2003)]%
        {bailenson2003interpersonal}
\bibfield{author}{\bibinfo{person}{Jeremy~N Bailenson}, \bibinfo{person}{Jim Blascovich}, \bibinfo{person}{Andrew~C Beall}, {and} \bibinfo{person}{Jack~M Loomis}.} \bibinfo{year}{2003}\natexlab{}.
\newblock \showarticletitle{Interpersonal distance in immersive virtual environments}.
\newblock \bibinfo{journal}{\emph{Personality and social psychology bulletin}} \bibinfo{volume}{29}, \bibinfo{number}{7} (\bibinfo{year}{2003}), \bibinfo{pages}{819--833}.
\newblock


\bibitem[Balk et~al\mbox{.}(2013)]%
        {balk2013simulator}
\bibfield{author}{\bibinfo{person}{Stacy~A Balk}, \bibinfo{person}{Dakota~B Bertola}, {and} \bibinfo{person}{Vaughan~W Inman}.} \bibinfo{year}{2013}\natexlab{}.
\newblock \showarticletitle{Simulator sickness questionnaire: twenty years later}. In \bibinfo{booktitle}{\emph{Driving Assessment Conference}}, Vol.~\bibinfo{volume}{7}. University of Iowa.
\newblock


\bibitem[Banholzer et~al\mbox{.}(2014)]%
        {banholzer2014impact}
\bibfield{author}{\bibinfo{person}{Sandra Banholzer}, \bibinfo{person}{James Kossin}, {and} \bibinfo{person}{Simon Donner}.} \bibinfo{year}{2014}\natexlab{}.
\newblock \showarticletitle{The impact of climate change on natural disasters}.
\newblock \bibinfo{journal}{\emph{Reducing disaster: Early warning systems for climate change}} (\bibinfo{year}{2014}), \bibinfo{pages}{21--49}.
\newblock


\bibitem[Basher(2006)]%
        {basher2006global}
\bibfield{author}{\bibinfo{person}{Reid Basher}.} \bibinfo{year}{2006}\natexlab{}.
\newblock \showarticletitle{Global early warning systems for natural hazards: systematic and people-centred}.
\newblock \bibinfo{journal}{\emph{Philosophical transactions of the royal society a: mathematical, physical and engineering sciences}} \bibinfo{volume}{364}, \bibinfo{number}{1845} (\bibinfo{year}{2006}), \bibinfo{pages}{2167--2182}.
\newblock


\bibitem[Blackie and Howland(1999)]%
        {blackie1999extension}
\bibfield{author}{\bibinfo{person}{CA Blackie} {and} \bibinfo{person}{HC Howland}.} \bibinfo{year}{1999}\natexlab{}.
\newblock \showarticletitle{An extension of an accommodation and convergence model of emmetropization to include the effects of illumination intensity}.
\newblock \bibinfo{journal}{\emph{Ophthalmic and Physiological Optics}} \bibinfo{volume}{19}, \bibinfo{number}{2} (\bibinfo{year}{1999}), \bibinfo{pages}{112--125}.
\newblock


\bibitem[Bubeck et~al\mbox{.}(2012)]%
        {bubeck2012review}
\bibfield{author}{\bibinfo{person}{Philip Bubeck}, \bibinfo{person}{Willem Jan~Wouter Botzen}, {and} \bibinfo{person}{Jeroen~CJH Aerts}.} \bibinfo{year}{2012}\natexlab{}.
\newblock \showarticletitle{A review of risk perceptions and other factors that influence flood mitigation behavior}.
\newblock \bibinfo{journal}{\emph{Risk Analysis: An International Journal}} \bibinfo{volume}{32}, \bibinfo{number}{9} (\bibinfo{year}{2012}), \bibinfo{pages}{1481--1495}.
\newblock


\bibitem[Burdea and Coiffet(2003)]%
        {burdea2003vr}
\bibfield{author}{\bibinfo{person}{Grigore Burdea} {and} \bibinfo{person}{Philippe Coiffet}.} \bibinfo{year}{2003}\natexlab{}.
\newblock \showarticletitle{Virtual Reality Technology}.
\newblock \bibinfo{journal}{\emph{Presence}}  \bibinfo{volume}{12} (\bibinfo{date}{12} \bibinfo{year}{2003}), \bibinfo{pages}{663--664}.
\newblock
\urldef\tempurl%
\url{https://doi.org/10.1162/105474603322955950}
\showDOI{\tempurl}


\bibitem[Chiba et~al\mbox{.}(2017)]%
        {chiba2017climate}
\bibfield{author}{\bibinfo{person}{Yohei Chiba}, \bibinfo{person}{Rajib Shaw}, {and} \bibinfo{person}{Sivapuram Prabhakar}.} \bibinfo{year}{2017}\natexlab{}.
\newblock \showarticletitle{Climate change-related non-economic loss and damage in Bangladesh and Japan}.
\newblock \bibinfo{journal}{\emph{International Journal of Climate Change Strategies and Management}} \bibinfo{volume}{9}, \bibinfo{number}{2} (\bibinfo{year}{2017}), \bibinfo{pages}{166--183}.
\newblock


\bibitem[Chiozzi et~al\mbox{.}(2023)]%
        {chiozzi2023natural}
\bibfield{author}{\bibinfo{person}{Andrea Chiozzi}, \bibinfo{person}{Elena Benvenuti}, {and} \bibinfo{person}{{\v{Z}}eljana Nikoli{\'c}}.} \bibinfo{year}{2023}\natexlab{}.
\newblock \bibinfo{booktitle}{\emph{Natural-Hazards Risk Assessment for Disaster Mitigation}}.
\newblock \bibinfo{publisher}{MDPI-Multidisciplinary Digital Publishing Institute}.
\newblock


\bibitem[Feldman et~al\mbox{.}(2016)]%
        {feldman2016communicating}
\bibfield{author}{\bibinfo{person}{David Feldman}, \bibinfo{person}{Santina Contreras}, \bibinfo{person}{Beth Karlin}, \bibinfo{person}{Victoria Basolo}, \bibinfo{person}{Richard Matthew}, \bibinfo{person}{Brett Sanders}, \bibinfo{person}{Douglas Houston}, \bibinfo{person}{Wing Cheung}, \bibinfo{person}{Kristen Goodrich}, \bibinfo{person}{Abigail Reyes}, {et~al\mbox{.}}} \bibinfo{year}{2016}\natexlab{}.
\newblock \showarticletitle{Communicating flood risk: Looking back and forward at traditional and social media outlets}.
\newblock \bibinfo{journal}{\emph{International Journal of Disaster Risk Reduction}}  \bibinfo{volume}{15} (\bibinfo{year}{2016}), \bibinfo{pages}{43--51}.
\newblock


\bibitem[Gagliardi et~al\mbox{.}(2023)]%
        {GAGLIARDI2023106141}
\bibfield{author}{\bibinfo{person}{Emanuele Gagliardi}, \bibinfo{person}{Gabriele Bernardini}, \bibinfo{person}{Enrico Quagliarini}, \bibinfo{person}{Michael Schumacher}, {and} \bibinfo{person}{Davide Calvaresi}.} \bibinfo{year}{2023}\natexlab{}.
\newblock \showarticletitle{Characterization and future perspectives of Virtual Reality Evacuation Drills for safe built environments: A Systematic Literature Review}.
\newblock \bibinfo{journal}{\emph{Safety Science}}  \bibinfo{volume}{163} (\bibinfo{year}{2023}), \bibinfo{pages}{106141}.
\newblock
\showISSN{0925-7535}
\urldef\tempurl%
\url{https://doi.org/10.1016/j.ssci.2023.106141}
\showDOI{\tempurl}


\bibitem[Gavas et~al\mbox{.}(2017)]%
        {gavas2017estimation}
\bibfield{author}{\bibinfo{person}{Rahul Gavas}, \bibinfo{person}{Debatri Chatterjee}, {and} \bibinfo{person}{Aniruddha Sinha}.} \bibinfo{year}{2017}\natexlab{}.
\newblock \showarticletitle{Estimation of cognitive load based on the pupil size dilation}. In \bibinfo{booktitle}{\emph{2017 IEEE international conference on systems, man, and cybernetics (SMC)}}. IEEE, \bibinfo{pages}{1499--1504}.
\newblock


\bibitem[Gibbs et~al\mbox{.}(2022)]%
        {GIBBS2022102717}
\bibfield{author}{\bibinfo{person}{Janet~K. Gibbs}, \bibinfo{person}{Marco Gillies}, {and} \bibinfo{person}{Xueni Pan}.} \bibinfo{year}{2022}\natexlab{}.
\newblock \showarticletitle{A comparison of the effects of haptic and visual feedback on presence in virtual reality}.
\newblock \bibinfo{journal}{\emph{International Journal of Human-Computer Studies}}  \bibinfo{volume}{157} (\bibinfo{year}{2022}), \bibinfo{pages}{102717}.
\newblock
\showISSN{1071-5819}
\urldef\tempurl%
\url{https://doi.org/10.1016/j.ijhcs.2021.102717}
\showDOI{\tempurl}


\bibitem[Hart(2006)]%
        {hart2006nasa}
\bibfield{author}{\bibinfo{person}{Sandra~G Hart}.} \bibinfo{year}{2006}\natexlab{}.
\newblock \showarticletitle{NASA-task load index (NASA-TLX); 20 years later}. In \bibinfo{booktitle}{\emph{Proceedings of the human factors and ergonomics society annual meeting}}, Vol.~\bibinfo{volume}{50}. Sage publications Sage CA: Los Angeles, CA, \bibinfo{pages}{904--908}.
\newblock


\bibitem[Hart and Staveland(1988)]%
        {hart1988development}
\bibfield{author}{\bibinfo{person}{Sandra~G Hart} {and} \bibinfo{person}{Lowell~E Staveland}.} \bibinfo{year}{1988}\natexlab{}.
\newblock \showarticletitle{Development of NASA-TLX (Task Load Index): Results of empirical and theoretical research}.
\newblock In \bibinfo{booktitle}{\emph{Advances in psychology}}. Vol.~\bibinfo{volume}{52}. \bibinfo{publisher}{Elsevier}, \bibinfo{pages}{139--183}.
\newblock


\bibitem[Herbert et~al\mbox{.}(2023)]%
        {herbert2023improving}
\bibfield{author}{\bibinfo{person}{Natalie Herbert}, \bibinfo{person}{Caroline Beckman}, \bibinfo{person}{Cade Cannedy}, \bibinfo{person}{Jinpu Cao}, \bibinfo{person}{Seung-Hyun Cho}, \bibinfo{person}{Stephanie Fischer}, \bibinfo{person}{ShihMing Huang}, \bibinfo{person}{Samantha~J Kramer}, \bibinfo{person}{Ortensia Lopez}, \bibinfo{person}{Sergio~Sanchez Lopez}, {et~al\mbox{.}}} \bibinfo{year}{2023}\natexlab{}.
\newblock \showarticletitle{Improving adaptation to wildfire smoke and extreme heat in frontline communities: evidence from a community-engaged pilot study in the San Francisco Bay Area}.
\newblock \bibinfo{journal}{\emph{Environmental Research Letters}} \bibinfo{volume}{18}, \bibinfo{number}{7} (\bibinfo{year}{2023}), \bibinfo{pages}{074026}.
\newblock


\bibitem[H{\"o}ppner et~al\mbox{.}(2010)]%
        {hoppner2010risk}
\bibfield{author}{\bibinfo{person}{Corina H{\"o}ppner}, \bibinfo{person}{Matthias Buchecker}, {and} \bibinfo{person}{Michael Br{\"u}ndl}.} \bibinfo{year}{2010}\natexlab{}.
\newblock \showarticletitle{Risk communication and natural hazards}.
\newblock \bibinfo{journal}{\emph{CapHaz-Net WP5 report}} (\bibinfo{year}{2010}).
\newblock


\bibitem[Houston et~al\mbox{.}(2015)]%
        {houston2015social}
\bibfield{author}{\bibinfo{person}{J~Brian Houston}, \bibinfo{person}{Joshua Hawthorne}, \bibinfo{person}{Mildred~F Perreault}, \bibinfo{person}{Eun~Hae Park}, \bibinfo{person}{Marlo Goldstein~Hode}, \bibinfo{person}{Michael~R Halliwell}, \bibinfo{person}{Sarah~E Turner~McGowen}, \bibinfo{person}{Rachel Davis}, \bibinfo{person}{Shivani Vaid}, \bibinfo{person}{Jonathan~A McElderry}, {et~al\mbox{.}}} \bibinfo{year}{2015}\natexlab{}.
\newblock \showarticletitle{Social media and disasters: a functional framework for social media use in disaster planning, response, and research}.
\newblock \bibinfo{journal}{\emph{Disasters}} \bibinfo{volume}{39}, \bibinfo{number}{1} (\bibinfo{year}{2015}), \bibinfo{pages}{1--22}.
\newblock


\bibitem[Hsu et~al\mbox{.}(2013)]%
        {5e941ffe32bc4cdebb72fcf9cdf55598}
\bibfield{author}{\bibinfo{person}{{Edbert B.} Hsu}, \bibinfo{person}{Yang Li}, \bibinfo{person}{{Jamil D.} Bayram}, \bibinfo{person}{David Levinson}, \bibinfo{person}{Samuel Yang}, {and} \bibinfo{person}{Colleen Monahan}.} \bibinfo{year}{2013}\natexlab{}.
\newblock \showarticletitle{State of Virtual Reality Based Disaster Preparedness and Response Training}.
\newblock \bibinfo{journal}{\emph{PLoS Currents}} \bibinfo{number}{APR 2013} (\bibinfo{year}{2013}).
\newblock
\showISSN{2157-3999}
\urldef\tempurl%
\url{https://doi.org/10.1371/currents.dis.1ea2b2e71237d5337fa53982a38b2aff}
\showDOI{\tempurl}


\bibitem[Irby et~al\mbox{.}(2009)]%
        {irby2009improving}
\bibfield{author}{\bibinfo{person}{Derek Irby}, \bibinfo{person}{Mahnas Mohammadi-Aragh}, \bibinfo{person}{Robert Moorhead}, {and} \bibinfo{person}{Phil Amburn}.} \bibinfo{year}{2009}\natexlab{}.
\newblock \showarticletitle{Improving the Understanding of Hurricanes: Visualizing Storm Surge}.
\newblock \bibinfo{journal}{\emph{Oceans}}  \bibinfo{volume}{1-3}, \bibinfo{pages}{1 -- 4}.
\newblock
\urldef\tempurl%
\url{https://doi.org/10.23919/OCEANS.2009.5422148}
\showDOI{\tempurl}


\bibitem[Israr et~al\mbox{.}(2014)]%
        {israr2014feel}
\bibfield{author}{\bibinfo{person}{Ali Israr}, \bibinfo{person}{Siyan Zhao}, \bibinfo{person}{Kaitlyn Schwalje}, \bibinfo{person}{Roberta Klatzky}, {and} \bibinfo{person}{Jill Lehman}.} \bibinfo{year}{2014}\natexlab{}.
\newblock \showarticletitle{Feel Effects: Enriching Storytelling with Haptic Feedback}.
\newblock \bibinfo{journal}{\emph{ACM Transactions on Applied Perception}}  \bibinfo{volume}{11} (\bibinfo{date}{09} \bibinfo{year}{2014}).
\newblock
\urldef\tempurl%
\url{https://doi.org/10.1145/2641570}
\showDOI{\tempurl}


\bibitem[Jiang et~al\mbox{.}(2022)]%
        {jiang2022scientometric}
\bibfield{author}{\bibinfo{person}{Sheng Jiang}, \bibinfo{person}{Junwei Ma}, \bibinfo{person}{Zhiyang Liu}, {and} \bibinfo{person}{Haixiang Guo}.} \bibinfo{year}{2022}\natexlab{}.
\newblock \showarticletitle{Scientometric analysis of artificial intelligence (AI) for geohazard research}.
\newblock \bibinfo{journal}{\emph{Sensors}} \bibinfo{volume}{22}, \bibinfo{number}{20} (\bibinfo{year}{2022}), \bibinfo{pages}{7814}.
\newblock


\bibitem[Johnston et~al\mbox{.}(2022)]%
        {johnston2022engaging}
\bibfield{author}{\bibinfo{person}{Kim~A Johnston}, \bibinfo{person}{Maureen Taylor}, {and} \bibinfo{person}{Barbara Ryan}.} \bibinfo{year}{2022}\natexlab{}.
\newblock \showarticletitle{Engaging communities to prepare for natural hazards: a conceptual model}.
\newblock \bibinfo{journal}{\emph{Natural Hazards}} \bibinfo{volume}{112}, \bibinfo{number}{3} (\bibinfo{year}{2022}), \bibinfo{pages}{2831--2851}.
\newblock


\bibitem[Katsiokalis et~al\mbox{.}(2023)]%
        {katsiokalis2023gonature}
\bibfield{author}{\bibinfo{person}{Minas Katsiokalis}, \bibinfo{person}{Elisavet Tsekeri}, \bibinfo{person}{Aikaterini Lilli}, \bibinfo{person}{Konstantinos Gobakis}, \bibinfo{person}{Dionysia Kolokotsa}, {and} \bibinfo{person}{Katerina Mania}.} \bibinfo{year}{2023}\natexlab{}.
\newblock \showarticletitle{GoNature AR: Air Quality \& Noise Visualization Through a Multimodal and Interactive Augmented Reality Experience}. In \bibinfo{booktitle}{\emph{Proceedings of the 2023 ACM International Conference on Interactive Media Experiences}}. \bibinfo{pages}{366--369}.
\newblock


\bibitem[Keeney et~al\mbox{.}(2012)]%
        {keeney2012multi}
\bibfield{author}{\bibinfo{person}{Harold~‘Jim’ Keeney}, \bibinfo{person}{Steve Buan}, {and} \bibinfo{person}{Laura Diamond}.} \bibinfo{year}{2012}\natexlab{}.
\newblock \showarticletitle{Multi-hazard early warning system of the United States National Weather Service}.
\newblock \bibinfo{journal}{\emph{Institutional partnerships in multi-hazard early warning systems: A compilation of seven national good practices and guiding principles}} (\bibinfo{year}{2012}), \bibinfo{pages}{115--157}.
\newblock


\bibitem[Kelman(2020)]%
        {kelman2020disaster}
\bibfield{author}{\bibinfo{person}{Ilan Kelman}.} \bibinfo{year}{2020}\natexlab{}.
\newblock \bibinfo{booktitle}{\emph{Disaster by choice: How our actions turn natural hazards into catastrophes}}.
\newblock \bibinfo{publisher}{Oxford University Press}.
\newblock


\bibitem[Kennedy et~al\mbox{.}(1993)]%
        {kennedy1993simulator}
\bibfield{author}{\bibinfo{person}{Robert~S Kennedy}, \bibinfo{person}{Norman~E Lane}, \bibinfo{person}{Kevin~S Berbaum}, {and} \bibinfo{person}{Michael~G Lilienthal}.} \bibinfo{year}{1993}\natexlab{}.
\newblock \showarticletitle{Simulator sickness questionnaire: An enhanced method for quantifying simulator sickness}.
\newblock \bibinfo{journal}{\emph{The international journal of aviation psychology}} \bibinfo{volume}{3}, \bibinfo{number}{3} (\bibinfo{year}{1993}), \bibinfo{pages}{203--220}.
\newblock


\bibitem[Kim and Kang(2010)]%
        {kim2010communication}
\bibfield{author}{\bibinfo{person}{Yong-Chan Kim} {and} \bibinfo{person}{Jinae Kang}.} \bibinfo{year}{2010}\natexlab{}.
\newblock \showarticletitle{Communication, neighbourhood belonging and household hurricane preparedness}.
\newblock \bibinfo{journal}{\emph{Disasters}} \bibinfo{volume}{34}, \bibinfo{number}{2} (\bibinfo{year}{2010}), \bibinfo{pages}{470--488}.
\newblock


\bibitem[Krejtz et~al\mbox{.}(2018)]%
        {krejtz2018eye}
\bibfield{author}{\bibinfo{person}{Krzysztof Krejtz}, \bibinfo{person}{Andrew~T Duchowski}, \bibinfo{person}{Anna Niedzielska}, \bibinfo{person}{Cezary Biele}, {and} \bibinfo{person}{Izabela Krejtz}.} \bibinfo{year}{2018}\natexlab{}.
\newblock \showarticletitle{Eye tracking cognitive load using pupil diameter and microsaccades with fixed gaze}.
\newblock \bibinfo{journal}{\emph{PloS one}} \bibinfo{volume}{13}, \bibinfo{number}{9} (\bibinfo{year}{2018}), \bibinfo{pages}{e0203629}.
\newblock


\bibitem[Lee et~al\mbox{.}(2023)]%
        {lee2023emotional}
\bibfield{author}{\bibinfo{person}{Jiyoung Lee}, \bibinfo{person}{Michael Hameleers}, {and} \bibinfo{person}{Soo~Yun Shin}.} \bibinfo{year}{2023}\natexlab{}.
\newblock \showarticletitle{The emotional effects of multimodal disinformation: How multimodality, issue relevance, and anxiety affect misperceptions about the flu vaccine}.
\newblock \bibinfo{journal}{\emph{New Media \& Society}} (\bibinfo{year}{2023}), \bibinfo{pages}{14614448231153959}.
\newblock


\bibitem[Lee and Shin(2022)]%
        {lee2022something}
\bibfield{author}{\bibinfo{person}{Jiyoung Lee} {and} \bibinfo{person}{Soo~Yun Shin}.} \bibinfo{year}{2022}\natexlab{}.
\newblock \showarticletitle{Something that they never said: multimodal disinformation and source vividness in understanding the power of AI-enabled deepfake news}.
\newblock \bibinfo{journal}{\emph{Media Psychology}} \bibinfo{volume}{25}, \bibinfo{number}{4} (\bibinfo{year}{2022}), \bibinfo{pages}{531--546}.
\newblock


\bibitem[Lin et~al\mbox{.}(2021)]%
        {lin2021early}
\bibfield{author}{\bibinfo{person}{J-T Lin}, \bibinfo{person}{Diego Melgar}, \bibinfo{person}{Amanda~M Thomas}, {and} \bibinfo{person}{Jacob Searcy}.} \bibinfo{year}{2021}\natexlab{}.
\newblock \showarticletitle{Early warning for great earthquakes from characterization of crustal deformation patterns with deep learning}.
\newblock \bibinfo{journal}{\emph{Journal of Geophysical Research: Solid Earth}} \bibinfo{volume}{126}, \bibinfo{number}{10} (\bibinfo{year}{2021}), \bibinfo{pages}{e2021JB022703}.
\newblock


\bibitem[Lindell and Perry(2012)]%
        {lindell2012protective}
\bibfield{author}{\bibinfo{person}{Michael~K Lindell} {and} \bibinfo{person}{Ronald~W Perry}.} \bibinfo{year}{2012}\natexlab{}.
\newblock \showarticletitle{The protective action decision model: Theoretical modifications and additional evidence}.
\newblock \bibinfo{journal}{\emph{Risk Analysis: An International Journal}} \bibinfo{volume}{32}, \bibinfo{number}{4} (\bibinfo{year}{2012}), \bibinfo{pages}{616--632}.
\newblock


\bibitem[Lundgren and McMakin(2018)]%
        {lundgren2018risk}
\bibfield{author}{\bibinfo{person}{Regina~E Lundgren} {and} \bibinfo{person}{Andrea~H McMakin}.} \bibinfo{year}{2018}\natexlab{}.
\newblock \bibinfo{booktitle}{\emph{Risk communication: A handbook for communicating environmental, safety, and health risks}}.
\newblock \bibinfo{publisher}{John Wiley \& Sons}.
\newblock


\bibitem[Martin et~al\mbox{.}(2022)]%
        {martin2022multimodality}
\bibfield{author}{\bibinfo{person}{Daniel Martin}, \bibinfo{person}{Sandra Malpica}, \bibinfo{person}{Diego Gutierrez}, \bibinfo{person}{Belen Masia}, {and} \bibinfo{person}{Ana Serrano}.} \bibinfo{year}{2022}\natexlab{}.
\newblock \showarticletitle{Multimodality in VR: A survey}.
\newblock \bibinfo{journal}{\emph{ACM Computing Surveys (CSUR)}} \bibinfo{volume}{54}, \bibinfo{number}{10s} (\bibinfo{year}{2022}), \bibinfo{pages}{1--36}.
\newblock


\bibitem[Mayhorn and McLaughlin(2014)]%
        {mayhorn2014warning}
\bibfield{author}{\bibinfo{person}{Christopher~B Mayhorn} {and} \bibinfo{person}{Anne~Collins McLaughlin}.} \bibinfo{year}{2014}\natexlab{}.
\newblock \showarticletitle{Warning the world of extreme events: A global perspective on risk communication for natural and technological disaster}.
\newblock \bibinfo{journal}{\emph{Safety science}}  \bibinfo{volume}{61} (\bibinfo{year}{2014}), \bibinfo{pages}{43--50}.
\newblock


\bibitem[McGlade et~al\mbox{.}(2019)]%
        {mcglade2019global}
\bibfield{author}{\bibinfo{person}{J McGlade}, \bibinfo{person}{G Bankoff}, \bibinfo{person}{J Abrahams}, \bibinfo{person}{SJ Cooper-Knock}, \bibinfo{person}{F Cotecchia}, \bibinfo{person}{P Desanker}, \bibinfo{person}{W Erian}, \bibinfo{person}{E Gencer}, \bibinfo{person}{L Gibson}, \bibinfo{person}{S Girgin}, {et~al\mbox{.}}} \bibinfo{year}{2019}\natexlab{}.
\newblock \bibinfo{title}{Global assessment report on disaster risk reduction 2019}.
\newblock
\newblock


\bibitem[Meehan et~al\mbox{.}(2002)]%
        {meehan2002physiological}
\bibfield{author}{\bibinfo{person}{Michael Meehan}, \bibinfo{person}{Brent Insko}, \bibinfo{person}{Mary Whitton}, {and} \bibinfo{person}{Frederick~P Brooks~Jr}.} \bibinfo{year}{2002}\natexlab{}.
\newblock \showarticletitle{Physiological measures of presence in stressful virtual environments}.
\newblock \bibinfo{journal}{\emph{Acm transactions on graphics (tog)}} \bibinfo{volume}{21}, \bibinfo{number}{3} (\bibinfo{year}{2002}), \bibinfo{pages}{645--652}.
\newblock


\bibitem[Mitsuhara and Shishibori(2020)]%
        {mitsuhara2020comparative}
\bibfield{author}{\bibinfo{person}{Hiroyuki Mitsuhara} {and} \bibinfo{person}{Masami Shishibori}.} \bibinfo{year}{2020}\natexlab{}.
\newblock \showarticletitle{Comparative experiments on simulated tornado experience via virtual reality and augmented reality}.
\newblock \bibinfo{journal}{\emph{The Journal of Information and Systems in Education}} \bibinfo{volume}{19}, \bibinfo{number}{1} (\bibinfo{year}{2020}), \bibinfo{pages}{21--31}.
\newblock


\bibitem[Mol et~al\mbox{.}(2022)]%
        {mol2022after}
\bibfield{author}{\bibinfo{person}{Jantsje~M Mol}, \bibinfo{person}{WJ~Wouter Botzen}, {and} \bibinfo{person}{Julia~E Blasch}.} \bibinfo{year}{2022}\natexlab{}.
\newblock \showarticletitle{After the virtual flood: Risk perceptions and flood preparednessafter virtual reality risk communication}.
\newblock \bibinfo{journal}{\emph{Judgment and Decision Making}} \bibinfo{volume}{17}, \bibinfo{number}{1} (\bibinfo{year}{2022}), \bibinfo{pages}{189--214}.
\newblock


\bibitem[Molan et~al\mbox{.}(2022)]%
        {molan2022why}
\bibfield{author}{\bibinfo{person}{Safa Molan}, \bibinfo{person}{Delene Weber}, {and} \bibinfo{person}{Matin Kor}.} \bibinfo{year}{2022}\natexlab{}.
\newblock \showarticletitle{‘Why can't they just evacuate early’ –a study of the effect of a virtual reality experience on residents' intentions related to a wildfire risk}.
\newblock \bibinfo{journal}{\emph{International Journal of Disaster Risk Reduction}}  \bibinfo{volume}{81} (\bibinfo{date}{09} \bibinfo{year}{2022}), \bibinfo{pages}{103268}.
\newblock
\urldef\tempurl%
\url{https://doi.org/10.1016/j.ijdrr.2022.103268}
\showDOI{\tempurl}


\bibitem[NCEI(2023)]%
        {NCEI2023}
\bibfield{author}{\bibinfo{person}{NCEI}.} \bibinfo{year}{2023}\natexlab{}.
\newblock \bibinfo{title}{U.S. Billion-Dollar Weather and Climate Disasters.}
\newblock
\newblock
\urldef\tempurl%
\url{https://www.ncei.noaa.gov/news/national-climate-202312#:~:text=Billion%2DDollar%20Weather%20and%20Climate%20Disasters,-In%202023%2C%20the&text=These%20disasters%20included%3A%2017%20severe,for%202023%20was%20%2492.9%20billion.}
\showURL{%
\tempurl}


\bibitem[of~Sciences et~al\mbox{.}(2018)]%
        {national2018emergency}
\bibfield{author}{\bibinfo{person}{National~Academies of Sciences}, \bibinfo{person}{Division on Engineering}, \bibinfo{person}{Physical Sciences}, \bibinfo{person}{Computer Science}, \bibinfo{person}{Telecommunications Board}, \bibinfo{person}{Committee on~the Future~of Emergency~Alert}, \bibinfo{person}{Warning Systems}, {and} \bibinfo{person}{Research Directions}.} \bibinfo{year}{2018}\natexlab{}.
\newblock \showarticletitle{Emergency alert and warning systems: Current knowledge and future research directions}.
\newblock  (\bibinfo{year}{2018}).
\newblock


\bibitem[Ogie et~al\mbox{.}(2018)]%
        {ogie2018disaster}
\bibfield{author}{\bibinfo{person}{Robert Ogie}, \bibinfo{person}{Juan Castilla~Rho}, \bibinfo{person}{Rodney~J Clarke}, {and} \bibinfo{person}{Alison Moore}.} \bibinfo{year}{2018}\natexlab{}.
\newblock \showarticletitle{Disaster risk communication in culturally and linguistically diverse communities: the role of technology}. In \bibinfo{booktitle}{\emph{Proceedings}}, Vol.~\bibinfo{volume}{2}. MDPI, \bibinfo{pages}{1256}.
\newblock


\bibitem[Ortiz et~al\mbox{.}(2023)]%
        {ortiz2023developing}
\bibfield{author}{\bibinfo{person}{Guadalupe Ortiz}, \bibinfo{person}{Pablo Aznar-Crespo}, {and} \bibinfo{person}{Antonio Aledo}.} \bibinfo{year}{2023}\natexlab{}.
\newblock \showarticletitle{Developing and pilot-testing warning messages for risk communication in natural disasters}.
\newblock \bibinfo{journal}{\emph{Environment Systems and Decisions}} (\bibinfo{year}{2023}), \bibinfo{pages}{1--20}.
\newblock


\bibitem[Oyshi et~al\mbox{.}(2022)]%
        {oyshi2022floodvis}
\bibfield{author}{\bibinfo{person}{Marzan~Tasnim Oyshi}, \bibinfo{person}{Verena Maleska}, \bibinfo{person}{Jochen Schanze}, \bibinfo{person}{Franziskus Borrmann}, \bibinfo{person}{Raimund Dachselt}, {and} \bibinfo{person}{Stefan Gumhold}.} \bibinfo{year}{2022}\natexlab{}.
\newblock \showarticletitle{FloodVis: Visualization of Climate Ensemble Flood Projections in Virtual Reality}.
\newblock


\bibitem[Palinko et~al\mbox{.}(2010)]%
        {palinko2010estimating}
\bibfield{author}{\bibinfo{person}{Oskar Palinko}, \bibinfo{person}{Andrew~L Kun}, \bibinfo{person}{Alexander Shyrokov}, {and} \bibinfo{person}{Peter Heeman}.} \bibinfo{year}{2010}\natexlab{}.
\newblock \showarticletitle{Estimating cognitive load using remote eye tracking in a driving simulator}. In \bibinfo{booktitle}{\emph{Proceedings of the 2010 symposium on eye-tracking research \& applications}}. \bibinfo{pages}{141--144}.
\newblock


\bibitem[Paton(2007)]%
        {paton2007preparing}
\bibfield{author}{\bibinfo{person}{Douglas Paton}.} \bibinfo{year}{2007}\natexlab{}.
\newblock \showarticletitle{Preparing for natural hazards: the role of community trust}.
\newblock \bibinfo{journal}{\emph{Disaster Prevention and Management: An International Journal}} \bibinfo{volume}{16}, \bibinfo{number}{3} (\bibinfo{year}{2007}), \bibinfo{pages}{370--379}.
\newblock


\bibitem[Paton and McClure(2013)]%
        {paton2013preparing}
\bibfield{author}{\bibinfo{person}{Douglas Paton} {and} \bibinfo{person}{John McClure}.} \bibinfo{year}{2013}\natexlab{}.
\newblock \bibinfo{booktitle}{\emph{Preparing for Disaster: Building household and community capacity}}.
\newblock \bibinfo{publisher}{Charles C Thomas Publisher}.
\newblock


\bibitem[Peterson and Manton(2008)]%
        {peterson2008monitoring}
\bibfield{author}{\bibinfo{person}{Thomas~C Peterson} {and} \bibinfo{person}{Michael~J Manton}.} \bibinfo{year}{2008}\natexlab{}.
\newblock \showarticletitle{Monitoring changes in climate extremes: a tale of international collaboration}.
\newblock \bibinfo{journal}{\emph{Bulletin of the American Meteorological Society}} \bibinfo{volume}{89}, \bibinfo{number}{9} (\bibinfo{year}{2008}), \bibinfo{pages}{1266--1271}.
\newblock


\bibitem[Peysakhovich et~al\mbox{.}(2017)]%
        {peysakhovich2017impact}
\bibfield{author}{\bibinfo{person}{Vsevolod Peysakhovich}, \bibinfo{person}{Fran{\c{c}}ois Vachon}, {and} \bibinfo{person}{Fr{\'e}d{\'e}ric Dehais}.} \bibinfo{year}{2017}\natexlab{}.
\newblock \showarticletitle{The impact of luminance on tonic and phasic pupillary responses to sustained cognitive load}.
\newblock \bibinfo{journal}{\emph{International Journal of Psychophysiology}}  \bibinfo{volume}{112} (\bibinfo{year}{2017}), \bibinfo{pages}{40--45}.
\newblock


\bibitem[Pierre et~al\mbox{.}(2015)]%
        {pierre2015luminance}
\bibfield{author}{\bibinfo{person}{Fabien Pierre}, \bibinfo{person}{Jean-Fran{\c{c}}ois Aujol}, \bibinfo{person}{Aur{\'e}lie Bugeau}, {and} \bibinfo{person}{Vinh-Thong Ta}.} \bibinfo{year}{2015}\natexlab{}.
\newblock \showarticletitle{Luminance-hue specification in the RGB space}. In \bibinfo{booktitle}{\emph{Scale Space and Variational Methods in Computer Vision: 5th International Conference, SSVM 2015, L{\`e}ge-Cap Ferret, France, May 31-June 4, 2015, Proceedings 5}}. Springer, \bibinfo{pages}{413--424}.
\newblock


\bibitem[Poussin et~al\mbox{.}(2014)]%
        {poussin2014factors}
\bibfield{author}{\bibinfo{person}{Jennifer~K Poussin}, \bibinfo{person}{WJ~Wouter Botzen}, {and} \bibinfo{person}{Jeroen~CJH Aerts}.} \bibinfo{year}{2014}\natexlab{}.
\newblock \showarticletitle{Factors of influence on flood damage mitigation behaviour by households}.
\newblock \bibinfo{journal}{\emph{Environmental Science \& Policy}}  \bibinfo{volume}{40} (\bibinfo{year}{2014}), \bibinfo{pages}{69--77}.
\newblock


\bibitem[Preechayasomboon and Rombokas(2021)]%
        {Preechayasomboon2021HapletsFW}
\bibfield{author}{\bibinfo{person}{Pornthep Preechayasomboon} {and} \bibinfo{person}{Eric Rombokas}.} \bibinfo{year}{2021}\natexlab{}.
\newblock \showarticletitle{Haplets: Finger-Worn Wireless and Low-Encumbrance Vibrotactile Haptic Feedback for Virtual and Augmented Reality}. In \bibinfo{booktitle}{\emph{Frontiers in Virtual Reality}}.
\newblock
\urldef\tempurl%
\url{https://api.semanticscholar.org/CorpusID:237566166}
\showURL{%
\tempurl}


\bibitem[Rahmayati et~al\mbox{.}(2017)]%
        {rahmayati2017understanding}
\bibfield{author}{\bibinfo{person}{Yenny Rahmayati}, \bibinfo{person}{Matthew Parnell}, {and} \bibinfo{person}{Vivien Himmayani}.} \bibinfo{year}{2017}\natexlab{}.
\newblock \showarticletitle{Understanding community-led resilience: the Jakarta floods experience}.
\newblock \bibinfo{journal}{\emph{Australian Journal of Emergency Management, The}} \bibinfo{volume}{32}, \bibinfo{number}{4} (\bibinfo{year}{2017}), \bibinfo{pages}{58--66}.
\newblock


\bibitem[Ritter and Chambers(2022)]%
        {ritter2022three}
\bibfield{author}{\bibinfo{person}{K. Ritter} {and} \bibinfo{person}{Terrence Chambers}.} \bibinfo{year}{2022}\natexlab{}.
\newblock \showarticletitle{Three-dimensional modeled environments versus 360 degree panoramas for mobile virtual reality training}.
\newblock \bibinfo{journal}{\emph{Virtual Reality}}  \bibinfo{volume}{26} (\bibinfo{date}{06} \bibinfo{year}{2022}).
\newblock
\urldef\tempurl%
\url{https://doi.org/10.1007/s10055-021-00502-9}
\showDOI{\tempurl}


\bibitem[Riva et~al\mbox{.}(2019)]%
        {riva2019virtual}
\bibfield{author}{\bibinfo{person}{Giuseppe Riva}, \bibinfo{person}{Jos{\'e} Guti{\'e}rrez-Maldonado}, \bibinfo{person}{Antonios Dakanalis}, {and} \bibinfo{person}{Marta Ferrer-Garc{\'\i}a}.} \bibinfo{year}{2019}\natexlab{}.
\newblock \showarticletitle{Virtual reality in the assessment and treatment of weight-related disorders}.
\newblock \bibinfo{journal}{\emph{Virtual reality for psychological and neurocognitive interventions}} (\bibinfo{year}{2019}), \bibinfo{pages}{163--193}.
\newblock


\bibitem[Schwind et~al\mbox{.}(2019)]%
        {schwind2019using}
\bibfield{author}{\bibinfo{person}{Valentin Schwind}, \bibinfo{person}{Pascal Knierim}, \bibinfo{person}{Nico Haas}, {and} \bibinfo{person}{Niels Henze}.} \bibinfo{year}{2019}\natexlab{}.
\newblock \showarticletitle{Using presence questionnaires in virtual reality}. In \bibinfo{booktitle}{\emph{Proceedings of the 2019 CHI conference on human factors in computing systems}}. \bibinfo{pages}{1--12}.
\newblock


\bibitem[Sermet and Demir(2018)]%
        {sermet2018flood}
\bibfield{author}{\bibinfo{person}{Yusuf Sermet} {and} \bibinfo{person}{Ibrahim Demir}.} \bibinfo{year}{2018}\natexlab{}.
\newblock \showarticletitle{Flood Action VR: A Virtual Reality Framework for Disaster Awareness and Emergency Response Training}.
\newblock
\showISBNx{978-1-4503-6314-3}
\urldef\tempurl%
\url{https://doi.org/10.1145/3306214.3338550}
\showDOI{\tempurl}


\bibitem[Shi et~al\mbox{.}(2020a)]%
        {shi2020impact2}
\bibfield{author}{\bibinfo{person}{Yangming Shi}, \bibinfo{person}{Jing Du}, {and} \bibinfo{person}{Darrell~A Worthy}.} \bibinfo{year}{2020}\natexlab{a}.
\newblock \showarticletitle{The impact of engineering information formats on learning and execution of construction operations: A virtual reality pipe maintenance experiment}.
\newblock \bibinfo{journal}{\emph{Automation in construction}}  \bibinfo{volume}{119} (\bibinfo{year}{2020}), \bibinfo{pages}{103367}.
\newblock


\bibitem[Shi et~al\mbox{.}(2021)]%
        {SHI2021105231}
\bibfield{author}{\bibinfo{person}{Yangming Shi}, \bibinfo{person}{John Kang}, \bibinfo{person}{Pengxiang Xia}, \bibinfo{person}{Oshin Tyagi}, \bibinfo{person}{Ranjana~K. Mehta}, {and} \bibinfo{person}{Jing Du}.} \bibinfo{year}{2021}\natexlab{}.
\newblock \showarticletitle{Spatial knowledge and firefighters’ wayfinding performance: A virtual reality search and rescue experiment}.
\newblock \bibinfo{journal}{\emph{Safety Science}}  \bibinfo{volume}{139} (\bibinfo{year}{2021}), \bibinfo{pages}{105231}.
\newblock
\showISSN{0925-7535}
\urldef\tempurl%
\url{https://doi.org/10.1016/j.ssci.2021.105231}
\showDOI{\tempurl}


\bibitem[Shi et~al\mbox{.}(2020b)]%
        {shi2020impact}
\bibfield{author}{\bibinfo{person}{Yangming Shi}, \bibinfo{person}{Yingfei Zheng}, \bibinfo{person}{Jing Du}, \bibinfo{person}{Qi Zhu}, {and} \bibinfo{person}{Xin Liu}.} \bibinfo{year}{2020}\natexlab{b}.
\newblock \showarticletitle{The impact of engineering information complexity on working memory development of construction workers: An eye-tracking investigation}. In \bibinfo{booktitle}{\emph{Construction Research Congress 2020}}. American Society of Civil Engineers Reston, VA, \bibinfo{pages}{89--98}.
\newblock


\bibitem[Simpson et~al\mbox{.}(2022)]%
        {SIMPSON2022101764}
\bibfield{author}{\bibinfo{person}{Mark Simpson}, \bibinfo{person}{Lace Padilla}, \bibinfo{person}{Klaus Keller}, {and} \bibinfo{person}{Alexander Klippel}.} \bibinfo{year}{2022}\natexlab{}.
\newblock \showarticletitle{Immersive storm surge flooding: Scale and risk perception in virtual reality}.
\newblock \bibinfo{journal}{\emph{Journal of Environmental Psychology}}  \bibinfo{volume}{80} (\bibinfo{year}{2022}), \bibinfo{pages}{101764}.
\newblock
\showISSN{0272-4944}
\urldef\tempurl%
\url{https://doi.org/10.1016/j.jenvp.2022.101764}
\showDOI{\tempurl}


\bibitem[Slovic(1987)]%
        {slovic1987perception}
\bibfield{author}{\bibinfo{person}{Paul Slovic}.} \bibinfo{year}{1987}\natexlab{}.
\newblock \showarticletitle{Perception of risk}.
\newblock \bibinfo{journal}{\emph{science}} \bibinfo{volume}{236}, \bibinfo{number}{4799} (\bibinfo{year}{1987}), \bibinfo{pages}{280--285}.
\newblock


\bibitem[Spence et~al\mbox{.}(2012)]%
        {spence2012colavita}
\bibfield{author}{\bibinfo{person}{Charles Spence}, \bibinfo{person}{Cesare Parise}, {and} \bibinfo{person}{Yi-Chuan Chen}.} \bibinfo{year}{2012}\natexlab{}.
\newblock \showarticletitle{The Colavita visual dominance effect}.
\newblock \bibinfo{journal}{\emph{The neural bases of multisensory processes}} (\bibinfo{year}{2012}).
\newblock


\bibitem[Sundar(2008)]%
        {sundar2018heuristic}
\bibfield{author}{\bibinfo{person}{S.~Shyam Sundar}.} \bibinfo{year}{2008}\natexlab{}.
\newblock \bibinfo{booktitle}{\emph{The MAIN Model: A Heuristic Approach to Understanding Technology Effects on Credibility}}.
\newblock


\bibitem[Susmayadi et~al\mbox{.}(2014)]%
        {susmayadi2014sustainable}
\bibfield{author}{\bibinfo{person}{I~Made Susmayadi}, \bibinfo{person}{Hidehiko Kanagae}, \bibinfo{person}{Wignyo Adiyoso}, \bibinfo{person}{Emi~Dwi Suryanti}, {et~al\mbox{.}}} \bibinfo{year}{2014}\natexlab{}.
\newblock \showarticletitle{Sustainable disaster risk reduction through effective risk communication media in Parangtritis tourism area, Yogyakarta}.
\newblock \bibinfo{journal}{\emph{Procedia Environmental Sciences}}  \bibinfo{volume}{20} (\bibinfo{year}{2014}), \bibinfo{pages}{684--692}.
\newblock


\bibitem[Sutton et~al\mbox{.}(2008)]%
        {sutton2008backchannels}
\bibfield{author}{\bibinfo{person}{Jeannette~N Sutton}, \bibinfo{person}{Leysia Palen}, {and} \bibinfo{person}{Irina Shklovski}.} \bibinfo{year}{2008}\natexlab{}.
\newblock \showarticletitle{Backchannels on the front lines: Emergency uses of social media in the 2007 Southern California Wildfires}.
\newblock  (\bibinfo{year}{2008}).
\newblock


\bibitem[Van~der Wel and Van~Steenbergen(2018)]%
        {van2018pupil}
\bibfield{author}{\bibinfo{person}{Pauline Van~der Wel} {and} \bibinfo{person}{Henk Van~Steenbergen}.} \bibinfo{year}{2018}\natexlab{}.
\newblock \showarticletitle{Pupil dilation as an index of effort in cognitive control tasks: A review}.
\newblock \bibinfo{journal}{\emph{Psychonomic bulletin \& review}}  \bibinfo{volume}{25} (\bibinfo{year}{2018}), \bibinfo{pages}{2005--2015}.
\newblock


\bibitem[Verguts et~al\mbox{.}(2015)]%
        {verguts2015adaptive}
\bibfield{author}{\bibinfo{person}{Tom Verguts}, \bibinfo{person}{Eliana Vassena}, {and} \bibinfo{person}{Massimo Silvetti}.} \bibinfo{year}{2015}\natexlab{}.
\newblock \showarticletitle{Adaptive effort investment in cognitive and physical tasks: a neurocomputational model}.
\newblock \bibinfo{journal}{\emph{Frontiers in Behavioral Neuroscience}}  \bibinfo{volume}{9} (\bibinfo{year}{2015}), \bibinfo{pages}{57}.
\newblock


\bibitem[Wachinger et~al\mbox{.}(2013)]%
        {wachinger2013risk}
\bibfield{author}{\bibinfo{person}{Gisela Wachinger}, \bibinfo{person}{Ortwin Renn}, \bibinfo{person}{Chloe Begg}, {and} \bibinfo{person}{Christian Kuhlicke}.} \bibinfo{year}{2013}\natexlab{}.
\newblock \showarticletitle{The risk perception paradox—implications for governance and communication of natural hazards}.
\newblock \bibinfo{journal}{\emph{Risk analysis}} \bibinfo{volume}{33}, \bibinfo{number}{6} (\bibinfo{year}{2013}), \bibinfo{pages}{1049--1065}.
\newblock


\bibitem[Wickham(2016)]%
        {ggplot2}
\bibfield{author}{\bibinfo{person}{Hadley Wickham}.} \bibinfo{year}{2016}\natexlab{}.
\newblock \bibinfo{booktitle}{\emph{ggplot2: Elegant Graphics for Data Analysis}}.
\newblock \bibinfo{publisher}{Springer-Verlag New York}.
\newblock
\showISBNx{978-3-319-24277-4}
\urldef\tempurl%
\url{https://ggplot2.tidyverse.org}
\showURL{%
\tempurl}


\bibitem[Witmer and Singer(1998)]%
        {witmer1998measuring}
\bibfield{author}{\bibinfo{person}{Bob~G Witmer} {and} \bibinfo{person}{Michael~J Singer}.} \bibinfo{year}{1998}\natexlab{}.
\newblock \showarticletitle{Measuring presence in virtual environments: A presence questionnaire}.
\newblock \bibinfo{journal}{\emph{Presence}} \bibinfo{volume}{7}, \bibinfo{number}{3} (\bibinfo{year}{1998}), \bibinfo{pages}{225--240}.
\newblock


\bibitem[Wu et~al\mbox{.}(2020)]%
        {wu2020exploring}
\bibfield{author}{\bibinfo{person}{Fei Wu}, \bibinfo{person}{Jerald Thomas}, \bibinfo{person}{Shreyas Chinnola}, {and} \bibinfo{person}{Evan~Suma Rosenberg}.} \bibinfo{year}{2020}\natexlab{}.
\newblock \showarticletitle{Exploring communication modalities to support collaborative guidance in virtual reality}. In \bibinfo{booktitle}{\emph{2020 IEEE conference on virtual reality and 3d user interfaces abstracts and workshops (VRW)}}. IEEE, \bibinfo{pages}{79--86}.
\newblock


\bibitem[Xu et~al\mbox{.}(2024)]%
        {xu2024augmented}
\bibfield{author}{\bibinfo{person}{Fang Xu}, \bibinfo{person}{Tianyu Zhou}, \bibinfo{person}{Tri Nguyen}, {and} \bibinfo{person}{Jing Du}.} \bibinfo{year}{2024}\natexlab{}.
\newblock \showarticletitle{Augmented reality in team-based search and rescue: Exploring spatial perspectives for enhanced navigation and collaboration}.
\newblock \bibinfo{journal}{\emph{Safety Science}}  \bibinfo{volume}{176} (\bibinfo{year}{2024}), \bibinfo{pages}{106556}.
\newblock


\bibitem[Yamori(2020)]%
        {yamori2020disaster}
\bibfield{author}{\bibinfo{person}{Katsuya Yamori}.} \bibinfo{year}{2020}\natexlab{}.
\newblock \bibinfo{booktitle}{\emph{Disaster risk communication}}.
\newblock \bibinfo{publisher}{Springer}.
\newblock


\bibitem[Ye et~al\mbox{.}(2022)]%
        {ye2022cognitive}
\bibfield{author}{\bibinfo{person}{Yang Ye}, \bibinfo{person}{Yangming Shi}, \bibinfo{person}{Pengxiang Xia}, \bibinfo{person}{John Kang}, \bibinfo{person}{Oshin Tyagi}, \bibinfo{person}{Ranjana~K Mehta}, {and} \bibinfo{person}{Jing Du}.} \bibinfo{year}{2022}\natexlab{}.
\newblock \showarticletitle{Cognitive characteristics in firefighter wayfinding Tasks: An Eye-Tracking analysis}.
\newblock \bibinfo{journal}{\emph{Advanced Engineering Informatics}}  \bibinfo{volume}{53} (\bibinfo{year}{2022}), \bibinfo{pages}{101668}.
\newblock


\bibitem[Zekveld et~al\mbox{.}(2011)]%
        {zekveld2011cognitive}
\bibfield{author}{\bibinfo{person}{Adriana~A Zekveld}, \bibinfo{person}{Sophia~E Kramer}, {and} \bibinfo{person}{Joost~M Festen}.} \bibinfo{year}{2011}\natexlab{}.
\newblock \showarticletitle{Cognitive load during speech perception in noise: The influence of age, hearing loss, and cognition on the pupil response}.
\newblock \bibinfo{journal}{\emph{Ear and hearing}} \bibinfo{volume}{32}, \bibinfo{number}{4} (\bibinfo{year}{2011}), \bibinfo{pages}{498--510}.
\newblock


\bibitem[Zhou et~al\mbox{.}(2023)]%
        {zhou2023cognition}
\bibfield{author}{\bibinfo{person}{Tianyu Zhou}, \bibinfo{person}{Pengxiang Xia}, \bibinfo{person}{Qi Zhu}, {and} \bibinfo{person}{Jing Du}.} \bibinfo{year}{2023}\natexlab{}.
\newblock \showarticletitle{Cognition-driven navigation assistive system for emergency indoor wayfinding (CogDNA): Proof of concept and evidence}.
\newblock \bibinfo{journal}{\emph{Safety science}}  \bibinfo{volume}{162} (\bibinfo{year}{2023}), \bibinfo{pages}{106100}.
\newblock


\end{thebibliography}

\appendix

\end{document}